\documentclass[draftclsnofoot,onecolumn,11pt]{IEEEtran}

\usepackage{amsmath}
\usepackage{amssymb}
\usepackage{amsthm}
\usepackage{dsfont}
\usepackage{graphicx}
\usepackage{cases,cite}
\newtheorem{thm} {Theorem}
\newtheorem{lem} {Lemma}

\newtheorem{rem} {Remark}
\newtheorem{assumption} {Assumption}

\begin{document}
\title{On the Influence of Informed Agents on Learning and Adaptation over Networks}

\author{Sheng-Yuan Tu, \IEEEmembership{Student Member,~IEEE}
        and Ali H. Sayed, \IEEEmembership{Fellow,~IEEE}
\thanks{This work was supported in part by NSF grants CCF-1011918 and CCF-0942936.
A short conference version of this work appears in \cite{Tu11d}. The
authors are with the Department of Electrical Engineering,
University of California, Los Angeles (e-mail: shinetu@ee.ucla.edu;
sayed@ee.ucla.edu)}}

\maketitle

\begin{abstract}
Adaptive networks consist of a collection of agents with adaptation
and learning abilities. The agents interact with each other on a
local level and diffuse information across the network through their
collaborations. In this work, we consider two types of agents:
informed agents and uninformed agents. The former receive new data
regularly and perform consultation and in-network tasks, while the
latter do not collect data and only participate in the consultation
tasks. We examine the performance of adaptive networks as a function
of the proportion of informed agents and their distribution in
space. The results reveal some interesting and surprising trade-offs
between convergence rate and mean-square performance. In particular,
among other results, it is shown that the performance of adaptive
networks does not necessarily improve with a larger proportion of
informed agents. Instead, it is established that the larger the
proportion of informed agents is, the faster the convergence rate of
the network becomes albeit at the expense of some deterioration in
mean-square performance. The results further establish that
uninformed agents play an important role in determining the
steady-state performance of the network, and that it is preferable
to keep some of the highly connected agents uninformed. The
arguments reveal an important interplay among three factors: the
number and distribution of informed agents in the network, the
convergence rate of the learning process, and the estimation
accuracy in steady-state. Expressions that quantify these relations
are derived, and simulations are included to support the theoretical
findings. We further apply the results to two models that are widely
used to represent behavior over complex networks, namely, the
Erdos-Renyi and scale-free models.
\end{abstract}


\section{Introduction}
Adaptive networks consist of a collection of spatially distributed
nodes that are linked together through a connection topology and
that cooperate with each other through local interactions. Adaptive
networks are well-suited to perform decentralized information
processing and inference tasks \cite{Lopes08,Cattivelli10} and to
model complex and self-organized behavior encountered in biological
systems \cite{Camazine03,Couzin09}, such as fish joining together in
schools \cite{Tu11} and birds flying in formation
\cite{Cattivelli11}.

In previous works on adaptive networks
\cite{Lopes08,Cattivelli10,Tu11}, and in other related studies on
distributed and combination algorithms
\cite{Arenas05,Predd06,Barbarossa07,Aysal09,Takahashi10b,Khan10,Candido10,Xia11b,Kozat10,
Sardellitti10,Kar11,Hu11,Theodoridis11,Lorenzo11}, the agents are
usually assumed to be homogeneous in that they all have similar
processing capabilities and are able to have continuous access to
information and measurements. However, it is generally observed in
nature that the behavior of a biological network is often driven
more heavily by a small fraction of the agents as happens, for
example, with bees and fish \cite{Avitabile75,Seeley79,Reebs00}.
This phenomenon motivates us to study in this paper adaptive
networks where only a \emph{fraction} of the nodes are assumed to be
informed, while the remaining nodes are uninformed. Informed nodes
collect data regularly and perform in-network processing tasks,
while uninformed nodes only participate in consultation tasks in the
manner explained in the sequel.

We shall examine how the transient and steady-state behavior of the
network are dependent on its topology and on the \emph{proportion}
of the informed nodes and their distribution in space. The results
will reveal some interesting and surprising trade-offs between
convergence rate and mean-square performance. In particular, among
other results, the analysis will show that the performance of
adaptive networks does not necessarily improve with a larger
proportion of informed nodes. Instead, it is discovered that the
larger the proportion of informed nodes is, the faster the
convergence rate of the network becomes albeit at the expense of
some deterioration in mean-square performance. The results also
establish that uninformed nodes play an important role in
determining the steady-state performance of the network, and that it
is preferable to maintain some of the highly connected nodes
uninformed. The analysis in the paper reveals the important
interplay that exists among three factors: the number of informed
nodes in a network, the convergence rate of the learning process,
and the estimation accuracy. We shall further apply the results to
two topology models that are widely used in the complex network
literature \cite{Newman06}, namely, the Erdos-Renyi and scale-free
models.

To establish the aforementioned results, a detailed
mean-square-error analysis of the network behavior is pursued.
However, the difficulty of the analysis is compounded by the fact
that nodes interact with each other and, therefore, they influence
each other's learning process and performance. Nevertheless, for
sufficiently small step-sizes, we will be able to derive an
expression for the network's mean-square deviation (MSD). By
examining this expression, we will establish that the MSD is
influenced by the eigen-structure of two matrices: the covariance
matrix representing the data statistical profile and the combination
matrix representing the network topology. We then study the
eigen-structure of these matrices and derive useful approximate
expressions for their eigenvalues and eigenvectors. The expressions
are subsequently used to reveal that the network MSD can be
decomposed into two components. We study the behavior of each
component as a function of the proportion of informed nodes; both
components show important differences in their behavior. When the
components are added together, a picture emerges that shows how the
performance of the network depends on the proportion of informed
nodes in an manner that supports analytically the popular wisdom
that \emph{more information is not necessarily better}
\cite{Surowiecki05}.

The organization of the paper is as follows. In Sections II and III,
we review the diffusion adaptation strategy and establish conditions
for the mean and mean-square stability of the networks in the
presence of uninformed nodes. In Section IV, we introduce two
popular models from the complex network literature. In Section V, we
analyze in some detail the structure of the mean-square performance
of the networks and reveal the effect of the network topology and
node distribution on learning and adaptation. Simulation results
appear in Section V in support of the theoretical findings.

\section{Diffusion Adaptation Strategy}
Consider a collection of $N$ nodes distributed over a domain in
space. Two nodes are said to be neighbors if they can share
information. The set of neighbors of node $k$, including $k$ itself,
is called the neighborhood of $k$ and is denoted by $\mathcal{N}_k$.
The nodes would like to estimate some unknown column vector,
$w^\circ$, of size $M$. At every time instant, $i$, each node $k$ is
able to observe realizations $\{d_k(i),u_{k,i}\}$ of a scalar random
process $\boldsymbol{d}_{k}(i)$ and a $1\times M$ vector random
process $\boldsymbol{u}_{k,i}$ with a positive-definite covariance
matrix,
$R_{u,k}=\mathbb{E}\boldsymbol{u}_{k,i}^*\boldsymbol{u}_{k,i}>0$,
where $\mathbb{E}$ denotes the expectation operator. All vectors in
our treatment are column vectors with the exception of the
regression vector, $\boldsymbol{u}_{k,i}$, which is taken to be a
row vector for convenience of presentation. The random processes
$\{\boldsymbol{d}_{k}(i),\boldsymbol{u}_{k,i}\}$ are assumed to be
related to $w^\circ$ via a linear regression model of the form
\cite{Sayed08}:
\begin{equation} \label{eq2}
    \boldsymbol{d}_k(i) = \boldsymbol{u}_{k,i}w^\circ+\boldsymbol{v}_k(i)
\end{equation}
where $\boldsymbol{v}_k(i)$ is measurement noise with variance
$\sigma^2_{v,k}$ and assumed to be spatially and temporally
independent with
\begin{equation}\label{eq41}
    \mathbb{E}\boldsymbol{v}^*_k(i)\boldsymbol{v}_l(j)=
    \sigma^2_{v,k}\cdot\delta_{kl}\cdot\delta_{ij}
\end{equation}
in terms of the Kronecker delta function. The noise
$\boldsymbol{v}_k(i)$ is assumed to be independent of
$\boldsymbol{u}_{l,j}$ for all $l$ and $j$. The regression data
$\boldsymbol{u}_{k,i}$ is likewise assumed to be spatially and
temporally independent. All random processes are assumed to be zero
mean. Note that we use boldface letters to denote random quantities
and normal letters to denote their realizations or deterministic
quantities. Models of the form (\ref{eq2})-(\ref{eq41}) are useful
in capturing many situations of interest, such as estimating the
parameters of some underlying physical phenomenon, or tracking a
moving target by a collection of nodes, or estimating the location
of a nutrient source or predator in biological networks (see, e.g.,
\cite{Tu11,Cattivelli11}).

\begin{figure}
\centering
\includegraphics[width=16em]{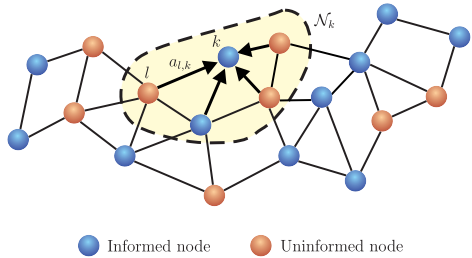}
\caption{A connected network with informed and uninformed nodes. The
weight $a_{l,k}$ scales the data transmitted from node $l$ to node
$k$ over the edge linking them.} \label{Fig_6}
\end{figure}

The objective of the network is to estimate $w^\circ$ in a
\emph{distributed} manner through an online learning process, where
each node is allowed to interact only with its neighbors. The nodes
estimate $w^\circ$ by seeking to minimize the following global cost
function:
\begin{equation} \label{eq1}
    J^{\text{glob}}(w)\triangleq\sum_{k=1}^{N}
    \mathbb{E}|\boldsymbol{d}_k(i)-\boldsymbol{u}_{k,i}w|^2
\end{equation}
Several diffusion adaptation schemes for solving (\ref{eq1}) in a
distributed manner were proposed in
\cite{Lopes08,Cattivelli10,Chen11}; the latter reference considers
more general cost functions. It was shown in these references,
through a constructive stochastic approximation and incremental
argument, that the structure of a near-optimal distributed solution
for (\ref{eq1}) takes the form of the Adapt-then-Combine (ATC)
strategy of \cite{Cattivelli10}; this strategy can be shown to
outperform other strategies in terms of mean-square performance
including consensus-based strategies \cite{Cattivelli10b,Tu12}.
Hence, we focus in this work on ATC updates. The ATC strategy
operates as follows. We select an $N\times N$ left-stochastic matrix
$A$ with nonnegative entries $\{a_{l,k}\geq 0\}$ satisfying:
\begin{equation} \label{eq50}
    A^T\mathds{1}=\mathds{1} \text{ and }
    a_{l,k}=0 \text{ if, and only if, $l\notin\mathcal{N}_k$}
\end{equation}
where $\mathds{1}$ is a vector of size $N$ with all entries equal to
one. The entry $a_{l,k}$ denotes the weight on the link connecting
node $l$ to node $k$, as shown in Fig. \ref{Fig_6}. Thus, condition
(\ref{eq50}) states that the weights on all links arriving at node
$k$ add up to one. Moreover, if two nodes $l$ and $k$ are linked,
then their corresponding entry $a_{l,k}$ is positive; otherwise,
$a_{l,k}$ is zero.  The ATC strategy consists of two steps. The
first step (\ref{eq49_1}) involves local adaptation, where node $k$
uses its own data $\{d_k(i),u_{k,i}\}$. This step updates the weight
estimate at node $k$ from $w_{k,i-1}$ to an intermediate value
$\psi_{k,i}$. The second step (\ref{eq49_2}) is a combination
(consultation) step where the intermediate estimates
$\{\psi_{l,i}\}$ from the neighborhood are combined through the
weights $\{a_{l,k}\}$ to obtain the updated weight estimate
$w_{k,i}$. The ATC strategy is described as follows:
\begin{subnumcases} {\label{eq49}}
    \psi_{k,i} = w_{k,i-1}+\mu_ku_{k,i}^*[d_k(i)-u_{k,i}w_{k,i-1}]\label{eq49_1}\\
    w_{k,i} = \sum_{l\in\mathcal{N}_k}a_{l,k}\psi_{l,i}
    \label{eq49_2}
\end{subnumcases}
where $\mu_k$ is the positive step-size used by node $k$. To model
uninformed nodes over the network, we shall set $\mu_k=0$ if node
$k$ is uninformed. We assume that the network contains at least one
informed node. In this model, uninformed nodes do not collect data
$\{d_{k}(i),u_{k,i}\}$ and, therefore, do not perform the adaptation
step (\ref{eq49_1}); they, however, continue to perform the
combination or consultation step (\ref{eq49_2}). In this way,
informed nodes have access to data and participate in the adaptation
and consultation steps, whereas uninformed nodes play an auxiliary
role through their participation in the consultation step only. This
participation is nevertheless important because it helps diffuse
information across the network. One of the main contributions of
this work is to examine how the proportion of informed nodes, and
how the spatial distribution of these informed nodes, influence the
learning and adaptation abilities of the network in terms of its
convergence rate and mean-square performance. It will follow from
the analysis that uninformed nodes also play an important role in
determining the network performance.

\section{Network Mean-Square Performance}
The mean-square performance of ATC networks has been studied in
detail in \cite{Cattivelli10} for the case where all nodes are
informed. Expressions for the network performance, and conditions
for its mean-square stability, were derived there by applying energy
conservation arguments \cite{Sayed08,Naffouri03}. In this section,
we start by showing how to extend the results to the case in which
only a fraction of the nodes are informed. The condition for
mean-square stability will need to be properly adjusted as explained
below in (\ref{eq6}) and (\ref{eq13}). We start by examining mean
stability.

\subsection{Mean Stability}
Let the error vector for any node $k$ be denoted by:
\begin{equation}
    \tilde{\boldsymbol{w}}_{k,i}\triangleq w^\circ-\boldsymbol{w}_{k,i}
\end{equation}
We collect all weight error vectors and step-sizes across the
network into a block vector and block matrix:
\begin{align}\label{eq84}
    \tilde{\boldsymbol{w}}_{i}\triangleq
    \text{col}\left\{\tilde{\boldsymbol{w}}_{1,i},\cdots,
    \tilde{\boldsymbol{w}}_{N,i}\right\},\quad
    \mathcal{M}\triangleq\text{diag}\{\mu_1I_M,\cdots,\mu_NI_M\}
\end{align}
where the notation $\text{col}\{\cdot\}$ denotes the vector that is
obtained by stacking its arguments on top of each other, and
$\text{diag}\{\cdot\}$ constructs a diagonal matrix from its
arguments. We also introduce the extended combination matrix:
\begin{equation}\label{eq96}
    \mathcal{A}\triangleq A\otimes I_M
\end{equation}
where the symbol $\otimes$ denotes the Kronecker product of two
matrices. Then, starting from (\ref{eq49_1})-(\ref{eq49_2}) and
using model (\ref{eq2}), some algebra will show that the global
error vector in (\ref{eq84}) evolves according to the recursion:
\begin{equation} \label{eq3}\boxed{
    \tilde{\boldsymbol{w}}_{i}=\mathcal{A}^T(I_{NM}-\mathcal{M} \boldsymbol{\mathcal{R}}_i)
    \tilde{\boldsymbol{w}}_{i-1}-\mathcal{A}^T\mathcal{M}
    \boldsymbol{s}_i}
\end{equation}
where
\begin{align}
    \boldsymbol{\mathcal{R}}_i\triangleq\text{diag}\{\boldsymbol{u}^*_{1,i}\boldsymbol{u}_{1,i},\cdots,
    \boldsymbol{u}^*_{N,i}\boldsymbol{u}_{N,i}\},\quad
    \boldsymbol{s}_i\triangleq\text{col}\{\boldsymbol{u}^*_{1,i}\boldsymbol{v}_{1,i},\cdots,
    \boldsymbol{u}^*_{N,i}\boldsymbol{v}_{N,i}\}
\end{align}
Since the regressors $\{\boldsymbol{u}_{k,i}\}$ are spatially and
temporally independent, then the $\{\boldsymbol{u}_{k,i}\}$ are
independent of $\tilde{\boldsymbol{w}}_{i-1}$. Taking expectation of
both sides of (\ref{eq3}), we find that the mean relation for
$\tilde{\boldsymbol{w}}_{i}$ evolves in time according to the
recursion:
\begin{equation}\label{eq5}\boxed{
    \mathbb{E}\tilde{\boldsymbol{w}}_{i}=\mathcal{B}\cdot
    \mathbb{E}\tilde{\boldsymbol{w}}_{i-1}}
\end{equation}
where we introduced the block matrices:
\begin{align}\label{eq9}
    \mathcal{B}\triangleq\mathcal{A}^T(I_{NM}-\mathcal{M}\mathcal{R}),\quad
    \mathcal{R}\triangleq \mathbb{E}\boldsymbol{\mathcal{R}}_{i}=\text{diag}\{R_{u,1},\cdots,R_{u,N}\}
\end{align}
In the following statement, we provide conditions to ensure mean
stability of the network, namely, that
$\mathbb{E}\tilde{\boldsymbol{w}}_{i}\rightarrow 0$ as
$i\rightarrow\infty$, even in the presence of uninformed nodes.

\begin{thm} [Mean stability]
The ATC network (\ref{eq49}) with at least one informed node
converges in the mean sense if the step-sizes $\{\mu_k\}$ and the
combination matrix $A$ satisfy the following two conditions:
\begin{enumerate}
\item For every informed node $l$, its step-size $\mu_l$
satisfies:
\begin{equation}\label{eq6}
    0<\mu_l\cdot\rho(R_{u,l})<2
\end{equation}
where the notation $\rho(\cdot)$ denotes the spectral radius of its
matrix argument.
\item There exists a finite
integer $j$ such that for every node $k$, there exists an informed
node $l$ satisfying:
\begin{equation} \label{eq13}
    \left[A^j\right]_{l,k}>0
\end{equation}
That is, the $(l,k)$th entry of $A^j$ is positive. [This condition
essentially ensures that there is a path from node $l$ to node $k$
in $j$ steps.]
\end{enumerate}
\end{thm}
{
\begin{proof}
We first introduce a block matrix norm. Let $\Sigma$ be an $N\times
N$ block matrix with blocks of size $M\times M$ each. Its block
matrix norm, $\|\Sigma\|_{b}$, is defined as:
\begin{equation}\label{eq44}
    \|\Sigma\|_{b}\triangleq\max_{1\leq k \leq N}
    \left(\sum_{l=1}^N \|\Sigma_{k,l} \|_2\right)
\end{equation}
where $\Sigma_{k,l}$ denotes the $(k,l)$th block of $\Sigma$ and
$\|\cdot\|_{2}$ denotes the largest singular value of its matrix
argument. That is, we first compute the 2-induced norm of each block
$\Sigma_{k,l}$ and then find the $\infty$-norm of the $N\times N$
matrix formed from the entries $\{\|\Sigma_{k,l}\|_2\}$. It can be
verified that (\ref{eq44}) satisfies the four conditions for a
matrix norm \cite{Horn85}. To prove mean stability of the ATC
network (\ref{eq49}), we need to show that conditions
(\ref{eq6})-(\ref{eq13}) guarantee $\rho(\mathcal{B})<1$, or
equivalently, $\rho(\mathcal{B}^j)<1$ for any $j$. Now, note that
\begin{equation} \label{eq14}
\begin{aligned}
    \rho(\mathcal{B}^j)\leq\|\mathcal{B}^j\|_{b}
    =\max_{1\leq k\leq N}
    \left(\sum_{l=1}^N\left\|\left[\mathcal{B}^j\right]_{k,l}\right\|_2\right)
\end{aligned}
\end{equation}
By the rules of matrix multiplication, the $(k,l)$th block (of size
$M\times M$) of the matrix $\mathcal{B}^j$ is given by:
\begin{equation}\label{eq12}
    \left[\mathcal{B}^j\right]_{k,l}=\sum_{m_1=1}^{N}\sum_{m_2=1}^{N}
    \cdots\sum_{m_{j-1}=1}^N
    \mathcal{B}_{k,m_1}\mathcal{B}_{m_1,m_2}\cdots\mathcal{B}_{m_{j-1},l}
\end{equation}
where $\mathcal{B}_{k,l}$ is the $(k,l)$th block (of size $M\times
M$) of the matrix $\mathcal{B}$ from (\ref{eq9}) and is given by
\begin{equation}\label{eq16}
    \mathcal{B}_{k,l}=a_{l,k}\cdot(I_M-\mu_lR_{u,l})
\end{equation}
Then, using the triangle inequality and the submultiplicative
property of norms, the 2-induced norm of $[\mathcal{B}^j]_{k,l}$ in
(\ref{eq12}) is bounded by:
\begin{equation}\label{eq15}
\begin{aligned}
    \left\|\left[\mathcal{B}^j\right]_{k,l}\right\|_2\leq
    \sum_{m_1=1}^{N}\sum_{m_2=1}^{N}
    \cdots\sum_{m_{j-1}=1}^N
    \|\mathcal{B}_{k,m_1}\|_2\cdot\|\mathcal{B}_{m_1,m_2}\|_2
    \cdots\|\mathcal{B}_{m_{j-1},l}\|_2
\end{aligned}
\end{equation}
Note that in the case where $l\in\mathcal{N}_m$, we obtain from
condition (\ref{eq6}) and expression (\ref{eq16}) that
\begin{equation}\label{eq76}
\begin{aligned}
    \|\mathcal{B}_{m,l}\|_2
    =a_{l,m}\cdot\rho\left(I_M-\mu_lR_{u,l}\right)
    \begin{cases}
    <a_{l,m}, &\text{if node $l$ is informed}\\
    =a_{l,m}, &\text{if node $l$ is uninformed}
    \end{cases}
\end{aligned}
\end{equation}
where we replaced the 2-induced norm with the spectral radius
because covariance matrices are Hermitian. Relation (\ref{eq76}) and
condition (\ref{eq50}) imply that
\begin{equation}\label{eq18}
\begin{aligned}
    \left\|\left[\mathcal{B}^j\right]_{k,l}\right\|_2\leq
    \sum_{m_1=1}^{N}\sum_{m_2=1}^{N}
    \cdots\sum_{m_{j-1}=1}^N
    a_{m_1,k}\cdot a_{m_2,m_1}\cdots a_{l,m_{j-1}}
\end{aligned}
\end{equation}
Strict inequality holds in (\ref{eq18}) if, and only if, the
sequence $(l,m_{j-1},\ldots,m_1,k)$ forms a path from node $l$ to
node $k$ using $j$ edges and there exists at least one informed node
along the path. Since we know from condition (\ref{eq13}) that there
is an informed node, say, node $l^\circ$, such that a path with $j$
edges exists from node $l^\circ$ to node $k$, we then get from
(\ref{eq14}) and (\ref{eq18}) that
\begin{equation}
\begin{aligned}
    \rho(\mathcal{B}^j)&\leq
    \max_{1\leq k\leq
    N}\left(\left\|\left[\mathcal{B}^j\right]_{k,l^\circ}\right\|_2+
    \sum_{l\neq l^\circ}\left\|\left[\mathcal{B}^j\right]_{k,l}\right\|_2\right)\\
    &<\max_{1\leq k\leq N}\sum_{l=1}^N
    \left(\sum_{m_1=1}^{N}\sum_{m_2=1}^{N}
    \cdots\sum_{m_{j-1}=1}^N
    a_{m_1,k}\cdot a_{m_2,m_1}\cdots a_{l,m_{j-1}}\right)\\
    &=\max_{1\leq k\leq N}\sum_{l=1}^N\left[A^j\right]_{l,k}\\
    &=1
\end{aligned}
\end{equation}
where the last equality is from condition (\ref{eq50}) because
$(A^T)^j\mathds{1}=\mathds{1}$ if $A^T\mathds{1}=\mathds{1}$.
\end{proof}}
Condition (\ref{eq13}) is satisfied if the matrix $A$ is primitive
\cite{Horn85}. Since, by (\ref{eq50}), $A$ is left-stochastic, it
follows from the Perron-Frobenius Theorem \cite{Horn85} that the
eigen-structure of $A$ satisfies certain prominent properties, which
will be useful in the sequel, namely, that (a) $A$ has an eigenvalue
at $\lambda=1$; (b) the eigenvalue at $\lambda=1$ has multiplicity
one; (c) all the entries of the right and left eigenvectors
associated with $\lambda=1$ are positive; and (d) $\rho(A)=1$ so
that all other eigenvalues of $A$ have magnitude strictly less than
one. We remark that since in this paper we will be dealing with
connected networks (where a path always exists between any two
arbitrary nodes), then condition (\ref{eq13}) is automatically
satisfied. As such, the ATC strategy (\ref{eq49}) will converge in
the mean whenever there exists at least one informed node with its
step-size satisfying condition (\ref{eq6}). In the next section, we
show that conditions (\ref{eq6})-(\ref{eq13}) further guarantee
mean-square convergence of the network when the step-sizes are
sufficiently small.

\subsection{Mean-Square Stability}
The network mean-square-deviation (MSD) is used to assess how well
the network estimates the weight vector, $w^\circ$. The MSD is
defined as follows:
\begin{equation}\label{eq85}
    \text{MSD}\triangleq\lim_{i\rightarrow\infty}
    \frac{1}{N}\sum_{k=1}^{N}\mathbb{E}\|\tilde{\boldsymbol{w}}_{k,i}\|^2
\end{equation}
where $\|\cdot\|$ denotes the Euclidean norm for vectors. To arrive
at an expression for the MSD, we first derive a variance relation
for the ATC network; the variance relation indicates how the
weighted mean-square error vector evolves over time \cite{Sayed08}.
Let $\Sigma$ denote an arbitrary nonnegative-definite Hermitian
matrix that we are free to choose, and let
$\sigma=\text{vec}(\Sigma)$ denote the vector that is obtained by
stacking the columns of $\Sigma$ on top of each other. We shall
interchangeably use the notation $\|x\|^2_{\Sigma}$ and
$\|x\|^2_{\sigma}$ to denote the same weighted square quantity,
$x^*\Sigma x$. Following the energy conservation approach of
\cite{Sayed08,Naffouri03}, we can motivate the following weighted
variance relation:
\begin{equation}\label{eq8}
\begin{aligned}
    \mathbb{E}\|\tilde{\boldsymbol{w}}_{i}\|^2_{\Sigma}=
    \mathbb{E}\left(\|\tilde{\boldsymbol{w}}_{i-1}\|^2_
    {(I_{NM}-\boldsymbol{\mathcal{R}}_i\mathcal{M} )\mathcal{A}\Sigma
    \mathcal{A}^T(I_{NM}-\mathcal{M} \boldsymbol{\mathcal{R}}_i)}\right)+
    \text{Tr}(\Sigma\mathcal{A}^T\mathcal{M} \mathcal{S}\mathcal{M}\mathcal{A})
\end{aligned}
\end{equation}
where
\begin{align}
    \mathcal{S}\triangleq \mathbb{E}\boldsymbol{s}_i\boldsymbol{s}_i^*
    =\text{diag}\{\sigma^2_{v,1}R_{u,1},
    \ldots,\sigma^2_{v,N}R_{u,N}\}
\end{align}
Relation (\ref{eq8}) can be derived from (\ref{eq3}) directly by
multiplying both sides from the left by
$\tilde{\boldsymbol{w}}^*_i\Sigma$ and taking expectations. Some
algebra will then show that for sufficiently small step-sizes,
expression (\ref{eq8}) can be approximated and rewritten as (see
\cite{Cattivelli10} for similar details, where terms that depend on
higher-order powers of the small step-sizes are ignored):
\begin{equation} \label{eq28}\boxed{
    \mathbb{E}\|\tilde{\boldsymbol{w}}_{i}\|^2_{\sigma}=
    \mathbb{E}\|\tilde{\boldsymbol{w}}_{i-1}\|^2_{F\sigma}+
    \left[\text{vec}(\mathcal{Y}^T)\right]^T\sigma}
\end{equation}
where
\begin{equation}\label{eq25}\boxed{
    \mathcal{F}\triangleq \mathcal{B}^T\otimes \mathcal{B}^*,\quad
    \mathcal{Y}\triangleq \mathcal{A}^T\mathcal{M}
    \mathcal{S}\mathcal{M}\mathcal{A}}
\end{equation}
Relation (\ref{eq28}) is very useful and it can be used to study the
transient behavior of the ATC network, as well as its steady-state
performance. The following result ensures that
$\mathbb{E}\|\tilde{\boldsymbol{w}}_{i}\|^2_{\sigma}$ remains
bounded and converges to some constant as $i$ goes to infinity.
\begin{thm}[Mean-square stability]
The ATC network (\ref{eq49}) with at least one informed node is
mean-square stable if the step-sizes $\{\mu_k\}$ and the combination
matrix $A$ satisfy conditions (\ref{eq6})-(\ref{eq13}), and the
step-sizes $\{\mu_k\}$ are sufficiently small such that higher-order
powers of them can be ignored.
\end{thm}
{
\begin{proof}
Expression (\ref{eq28}) holds for sufficiently small step-sizes. As
shown in \cite{Cattivelli10}, the mean-square convergence of
(\ref{eq28}) is guaranteed if $\rho\left(F\right)<1$. But since
\begin{equation}
\begin{aligned}
    \rho\left(\mathcal{F}\right)=\rho\left
    (\mathcal{B}^T \otimes \mathcal{B}^*\right)
    =\left[\rho\left(\mathcal{B}\right)\right]^2
\end{aligned}
\end{equation}
and conditions (\ref{eq6})-(\ref{eq13}) guarantee
$\rho\left(\mathcal{B}\right)<1$ from Theorem 1, it also holds that
$\rho\left(\mathcal{F}\right)<1$.
\end{proof}}

\subsection{Mean-Square Performance}
Now, assume the network is mean-square stable and let the time index
$i$ tend to infinity. From (\ref{eq28}), we obtain the steady-state
relation
\begin{equation}
    \lim_{i\rightarrow\infty}\mathbb{E}
    \|\tilde{\boldsymbol{w}}_i\|^2_{\left(I_{N^2M^2}-\mathcal{F}\right)\sigma}
    =\left[\text{vec}(\mathcal{Y}^T)\right]^T\sigma
\end{equation}
Since the eigenvalues of the matrix $\mathcal{F}$ are within the
unit disc, the matrix $\left(I_{N^2M^2}-\mathcal{F}\right)$ is
invertible. Thus, the network MSD, as given by (\ref{eq85}), can be
obtained by choosing
$\sigma=\left(I_{N^2M^2}-\mathcal{F}\right)^{-1}\text{vec}(I_{NM})/N$,
which leads to the following useful expression
\begin{equation} \label{eq4}\boxed{
\begin{aligned}
    \text{MSD}=\frac{1}{N}\left[\text{vec}(\mathcal{Y}^T)\right]^T
    \left(I_{N^2M^2}-\mathcal{F}\right)^{-1}\text{vec}(I_{NM})
\end{aligned}}
\end{equation}
Expression (\ref{eq4}) relates the network MSD to the quantities
$\{\mathcal{Y}, \mathcal{F}\}$ defined by (\ref{eq25}). These
quantities contain information about the data statistical profile,
the spatial distribution of informed nodes, and the network topology
through their dependence on $\{\mathcal{R}, \mathcal{M},
\mathcal{A}\}$. Using the following equalities for arbitrary
matrices $\{U,W,\Sigma\}$ of compatible dimensions:
\begin{align}\label{eq17}
    \text{vec}(U\Sigma W)=(W^T\otimes U)\sigma,\quad
    \text{Tr}(\Sigma W)=\left[\text{vec}(W^T)\right]^T\sigma
\end{align}
and the fact that, for any stable matrix $F$, it holds:
\begin{equation}
\begin{aligned}
    (I_{N^2M^2}-\mathcal{F})^{-1}&=\sum_{j=0}^\infty \mathcal{F}^j
\end{aligned}
\end{equation}
we can obtain an alternative expression for the network MSD from
(\ref{eq25}) and (\ref{eq4}), namely,
\begin{equation} \label{eq31}\boxed{
\begin{aligned}
    \text{MSD}=\frac{1}{N}\sum_{j=0}^\infty\text{Tr}
    [\mathcal{B}^j\mathcal{Y}(\mathcal{B}^*)^j]
\end{aligned}}
\end{equation}
This expression for the MSD will be the starting point for our
analysis further ahead, when we examine the influence of the
proportion of informed nodes on network performance.

\subsection{Convergence Rate}
We denote the convergence rate of the ATC strategy (\ref{eq49}) by
$r$ so that the smaller the value of $r$ is, the faster the rate of
convergence of $\mathbb{E}\|\tilde{\boldsymbol{w}}_{i}\|^2$ is
towards its steady-state value. As indicated by (\ref{eq28}), the
convergence rate is determined by the spectral radius of the matrix
$F$ in (\ref{eq25}), i.e.,
\begin{equation}\label{eq89}\boxed{
    r=\rho(\mathcal{F})=\left[\rho(\mathcal{B})\right]^2}
\end{equation}
Let $\mathcal{N}_I$ denote the set of informed nodes, i.e.,
$k\in\mathcal{N}_I$ if node $k$ is informed. From now on, we
introduce the assumption below, which essentially assumes that all
informed nodes have similar processing abilities in that they use
the same step-size value while observing processes arising from the
same statistical distribution.
\begin{assumption}
Assume that $\mu_k=\mu$ for all informed nodes and that the
covariance matrices across all nodes are also uniform, i.e.,
$R_{u,k}=R_u$. We continue to assume that the step-size is
sufficiently small so that it holds that $0<\mu\cdot\rho(R_u)< 1$.
\end{assumption}
\noindent Then, we have the following useful result.
\begin{lem} [Faster convergence rate] \label{lem_1}
Consider two configurations of the same network: one with
$\mathcal{N}_{I,1}$ informed nodes and another with
$\mathcal{N}_{I,2}$ informed nodes. Let $r_1$ and $r_2$ denote the
corresponding convergence rates for these two informed
configurations. If $\mathcal{N}_{I,2}\supseteq \mathcal{N}_{I,1}$,
then $r_2\leq r_1$.
\end{lem}
{
\begin{proof}
Under Assumption 1, we have that
\begin{equation}
\begin{aligned}
    I_M-\mu_{l}R_{u,l}
    =\begin{cases}
    I_M-\mu R_u, &\text{if node $l$ is informed}\\
    I_M, &\text{if node $l$ is uninformed}
    \end{cases}
\end{aligned}
\end{equation}
Then, the matrix $[\mathcal{B}^j]_{k,l}$ in (\ref{eq12}) can be
written as:
\begin{equation}\label{eq47}
\begin{aligned}
    \left[\mathcal{B}^j\right]_{k,l}=\sum_{m_1=1}^{N}\sum_{m_2=1}^{N}
    \cdots\sum_{m_{j-1}=1}^N
    a_{m_1,k}\cdot a_{m_2,m_1}\cdots a_{l,m_{j-1}}\cdot
    (I_M-\mu R_u)^{q_{l,k}}
\end{aligned}
\end{equation}
where the exponent $q_{l,k}$ denotes the number of informed nodes
along the path $(l,m_{j-1},\ldots,m_1,k)$. Note that
$[\mathcal{B}^j]_{k,l}$ is a nonnegative-definite matrix because
$(I_M-\mu R_u)>0$ in view of the condition $0<\mu\rho(R_u)<1$. In
fact, all eigenvalues of $(I_M-\mu R_u)$ lie within the line segment
$(0,1)$. Moreover, since $\mathcal{N}_{I,1}\subseteq
\mathcal{N}_{I,2}$, we have that $q^{(1)}_{l,k}\leq q^{(2)}_{l,k}$
and, therefore, the matrix difference
\begin{equation}
\begin{aligned}
    \left[\mathcal{B}^{(1)j}\right]_{k,l}-
    \left[\mathcal{B}^{(2)j}\right]_{k,l}
    =&\sum_{m_1=1}^{N}\sum_{m_2=1}^{N}
    \cdots\sum_{m_{j-1}=1}^N
    a_{m_1,k}\cdot a_{m_2,m_1}\cdots a_{l,m_{j-1}}\\
    &\times\left[\left(I-\mu
    R_u\right)^{q^{(1)}_{l,k}}-\left(I-\mu
    R_u\right)^{q^{(2)}_{l,k}}\right]
\end{aligned}
\end{equation}
is a nonnegative-definite matrix, where the superscripts denote the
indices of the informed configurations, $\mathcal{N}_{I,1}$ or
$\mathcal{N}_{I,2}$. Since $[\mathcal{B}^{(1)j}]_{k,l}$,
$[\mathcal{B}^{(2)j}]_{k,l}$, and
$[\mathcal{B}^{(1)j}]_{k,l}-[\mathcal{B}^{(2)j}]_{k,l}$ are all
nonnegative-definite, then it must hold that
\begin{equation} \label{eq35}
    \left\|\left[\mathcal{B}^{(1)j}\right]_{k,l}\right\|_2
    \geq \left\|\left[\mathcal{B}^{(2)j}\right]_{k,l}\right\|_2
\end{equation}
Relation (\ref{eq35}) can be established by contradiction. Suppose
that (\ref{eq35}) does not hold, i.e.,
$\rho([\mathcal{B}^{(1)j}]_{k,l})<\rho([\mathcal{B}^{(2)j}]_{k,l})$
as $[\mathcal{B}^{(1)j}]_{k,l}$ and $[\mathcal{B}^{(2)j}]_{k,l}$ are
Hermitian from (\ref{eq47}). In addition, let $x$ denote the
eigenvector that is associated with the largest eigenvalue of
$[\mathcal{B}^{(2)j}]_{k,l}$, i.e.,
$([\mathcal{B}^{(2)j}]_{k,l})x=\rho([\mathcal{B}^{(2)j}]_{k,l})x$.
Then, we obtain the following contradiction to the
nonnegative-definiteness of
$[\mathcal{B}^{(1)j}]_{k,l}-[\mathcal{B}^{(2)j}]_{k,l}$:
\begin{equation}\label{eq7}
    x^*\left(\left[\mathcal{B}^{(1)j}\right]_{k,l}-
    \left[\mathcal{B}^{(2)j}\right]_{k,l}\right)x=
    x^*\left(\left[\mathcal{B}^{(1)j}\right]_{k,l}\right)x-
    \rho\left(\left[\mathcal{B}^{(2)j}\right]_{k,l}\right)x^*x<0
\end{equation}
by the Rayleigh-Ritz Theorem \cite{Horn85}. By the definition of the
block matrix norm in (\ref{eq44}), we arrive at
\begin{equation}
    \left(\left\|\left[\mathcal{B}^{(1)j}\right]\right\|_b\right)^{1/j}
    \geq
    \left(\left\|\left[\mathcal{B}^{(2)j}\right]\right\|_b\right)^{1/j}
\end{equation}
for all $j$. Let $j$ tend to infinity and we obtain that
\begin{equation}
    \rho\left(\mathcal{B}^{(1)}\right)\geq\rho\left(\mathcal{B}^{(2)}\right)
\end{equation}
where we used the fact that
$\rho(\mathcal{B})=\lim_{j\rightarrow\infty}(\|\mathcal{B}^j\|)^{1/j}$
for any matrix norm \cite{Horn85}.
\end{proof}}
The result of Lemma \ref{lem_1} shows that if we enlarge the set of
informed nodes, the convergence rate decreases and convergence
becomes faster. The following result provides bounds for the
convergence rate.
\begin{lem}[Bound on convergence rate]\label{thm_1}
The convergence rate is bounded by
\begin{equation}\label{eq90}
    \left[1-\mu\cdot \lambda_M(R_{u})\right]^2 \leq r < 1
\end{equation}
where $\lambda_M(R_{u})$ denotes the smallest eigenvalue of $R_u$.
\end{lem}
{
\begin{proof}
Since the ATC network is mean-square stable, i.e.,
$\rho(\mathcal{B})<1$, the upper bound is obvious. On the other
hand, from Lemma \ref{lem_1}, the value of $\rho(\mathcal{B})$
achieves its minimum value when all nodes are informed, i.e., the
matrix $\mathcal{M}$ in (\ref{eq84}) becomes $\mathcal{M}=\mu
I_{NM}$. In this case, the matrix $\mathcal{B}$ in (\ref{eq9}) can
be written as:
\begin{equation}
    \mathcal{B}^\circ=A^T\otimes(I_M-\mu R_u)
\end{equation}
where the superscript is used to denote the matrix $\mathcal{B}$
when all nodes are informed. Then,
\begin{equation}\label{eq77}
\begin{aligned}
    \rho(\mathcal{B})&\geq \rho(\mathcal{B}^\circ)\\
    &=\rho(A^T)\cdot\rho(I_M-\mu R_u)
\end{aligned}
\end{equation}
We already know that $\rho(A^T)=1$. In addition, because $(I_M-\mu
R_u)>0$, we have that
\begin{equation}
    \rho(I_M-\mu R_u) =1-\mu\cdot \lambda_M(R_u)
\end{equation}
and we arrive at the lower bound in (\ref{eq90}).
\end{proof}}

\section{Two Network Topology Models}
Before examining the effect of informed nodes on network
performance, we pause to introduce two popular models that are
widely used in the study of complex networks. We shall call upon
these models later to illustrate the theoretical findings of the
article. For both models, we let $n_k$ denote the degree (number of
neighbors) of node $k$. Note that since node $k$ is a neighbor of
itself, we have $n_k\geq 1$. In addition, we assume the network
topology is symmetric so that if node $l$ is a neighbor of node $k$,
then node $k$ is also a neighbor of node $l$.

\subsection{Erdos-Renyi Model}
In the Erdos-Renyi model \cite{Erdos60}, there is a single parameter
called \emph{edge probability} and is denoted by $p\in[0,1]$. The
edge probability specifies the probability that two distinct nodes
are connected. In this way, the degree distribution of any node $k$
becomes a random variable and is distributed according to a binomial
distribution, i.e.,
\begin{equation}\label{eq33}
    f(n_k)=\begin{pmatrix}
    N-1 \\ n_k-1\end{pmatrix}
    p^{n_k-1}(1-p)^{N-n_k}
\end{equation}
The expected degree for node $k$, denoted by $\bar{n}_k$, is then
\begin{equation}\label{eq86}
    \bar{n}_k=(N-1)p+1
\end{equation}
Note that, in this model, all nodes have the same expected degree
since the right-hand side is independent of $k$. Therefore, the
expected network degree, $\bar{\eta}$, becomes
\begin{equation}\label{eq88}
    \bar{\eta}\triangleq\frac{1}{N}\sum_{k=1}^N\bar{n}_k
    =(N-1)p+1
\end{equation}

\subsection{Scale-Free Model}
The Erdos-Renyi model does not capture several prominent features of
real networks such as the \emph{small-world phenomenon} and the
\emph{power-law degree distribution} \cite{Newman06}. The
small-world phenomenon refers to the fact that the number of edges
between two arbitrary nodes is small on average. The power-law
degree distribution refers to the fact that the number of nodes with
degree $n_k$ falls off as an inverse power of $n_k$, namely,
\begin{equation} \label{eq19}
    f(n_k)\sim cn_k^{-\gamma}
\end{equation}
with two positive constants $c$ and $\gamma$. Networks with degree
distributions of the form (\ref{eq19}) are called scale-free
networks \cite{Barabasi03} and can be generated using preferential
attachment models. We briefly describe the model proposed by
\cite{Barabasi99}. The model starts with a small connected network
with $N_0$ nodes. At every iteration, we add a new node, which will
connect to $m\leq N_0$ distinct nodes besides itself. The
probability of connecting to a node is proportional to its degree.
As time evolves, nodes with higher degree are more likely to be
connected to new nodes. Eventually, there are a few nodes that
connect to most of the network. This phenomenon is observed in real
networks, such as the Internet \cite{Newman06}. If $N\gg N_0$, the
expected degree of the network approximates to
\begin{equation}\label{eq70}
    \bar{\eta}\approx 2m+1
\end{equation}
because every new arrival node contributes $2m+1$ degrees to the
network.

\section{Effect of Topology and Node Distribution}
We are now ready to examine in some detail the effect of network
topology and node distribution on the behavior of the network MSD
given by (\ref{eq31}) and the convergence rate given by
(\ref{eq89}).

\subsection{Eigen-structure of $\mathcal{B}$}
To begin with, we observe from (\ref{eq31}) and (\ref{eq89}) that
the network MSD and convergence rate depend on the matrix
$\mathcal{B}$ from (\ref{eq9}) in a non-trivial manner. To gain
insight into the network performance, we need to examine closely the
eigen-structure of $\mathcal{B}$, which is related to the
combination matrix $A$ and the covariance matrix $R_u$. We start
from the eigen-structure of $A$. To facilitate the analysis, we
assume that $A$ is diagonalizable, i.e., there exists an invertible
matrix, $U$, and a diagonal matrix, $\Lambda$, such that
\begin{equation}\label{eq67}
    A^T=U\Lambda U^{-1}
\end{equation}
Now, let $r_k$ and $s_k$ ($k=1,\ldots,N$) denote an arbitrary pair
of right and left eigenvectors of $A^T$ corresponding to the
eigenvalue $\lambda_k(A)$. Then,
\begin{align}\label{eq92}
    U = \begin{bmatrix}r_1 & \cdots & r_N\end{bmatrix},\quad
    U^{-1} = \text{col}\{s^*_1,\ldots,s^*_N\},\quad
    \Lambda = \text{diag}\{\lambda_1(A),\ldots,\lambda_N(A)\}
\end{align}
Obviously, it holds that $s^*_lr_k=\delta_{kl}$ since $UU^{-1}=I_N$.
We further assume that the right eigenvectors of $A^T$ satisfy:
\begin{equation}\label{eq10}
    |r^*_lr_k| \ll \|r_k\|^2
\end{equation}
for $l\neq k$. Condition (\ref{eq10}) states that the $\{r_k\}$ are
approximately orthogonal (see example below). Without loss of
generality, we order the eigenvalues of $A^T$ in decreasing order
and assume $1=\lambda_1(A)>|\lambda_2(A)|\geq\cdots
\geq|\lambda_N(A)|$. The eigen-decomposition of $A^T$ can also be
written as:
\begin{equation}\label{eq60}\boxed{
    A^T=\sum_{k=1}^N\lambda_k(A)\cdot r_ks_k^*}
\end{equation}
Note that any symmetric combination matrix satisfies both conditions
(\ref{eq67}) and (\ref{eq10}) since then $r^*_lr_k=\delta_{kl}$.
Another example of a useful combination matrix $A$ that is not
symmetric but still satisfies (\ref{eq67}) is the uniform
combination matrix, i.e.,
\begin{equation}\label{eq21}
    a_{l,k}=
    \begin{cases}
    1/n_k, &\text{if $l\in\mathcal{N}_k$}\\
    0, &\text{otherwise}
    \end{cases}
\end{equation}
\begin{lem}[Diagonalization of uniform combination matrix] \label{lem_2}
For a connected and symmetric network graph, the matrix $A$ defined
by (\ref{eq21}) is diagonalizable and has real eigenvalues.
\end{lem}
{
\begin{proof}
We introduce the degree matrix, $D$, and the adjacency matrix, $C$,
of the network graph, whose entries are defined as follows:
\begin{align}\label{eq65}
    D=\text{diag}\{n_1,\ldots,n_N\},\quad
    [C]_{k,l}=\begin{cases}
    1, &\text{if $l\in\mathcal{N}_k$}\\
    0, &\text{otherwise}
    \end{cases}
\end{align}
Then, it is straightforward to verify that the matrix $A^T$ in
(\ref{eq21}) can be written as:
\begin{equation}
    A^T = D^{-1}C
\end{equation}
which shows that $A^T$ is similar to the real-valued matrix $A_s$
defined by:
\begin{equation}\label{eq64}
\begin{aligned}
    A_s &\triangleq D^{1/2}A^TD^{-1/2}\\
    &= D^{-1/2}CD^{-1/2}
\end{aligned}
\end{equation}
where $D^{1/2}=\text{diag}\{\sqrt{n_1},\ldots,\sqrt{n_N}\}$. Since
the topology is assumed to be symmetric, the matrix $C$ is
symmetric, and so is $A_s$. Therefore, there exists an orthogonal
matrix, $U_s$, and a diagonal matrix with real diagonal entries,
$\Lambda$, such that
\begin{equation}
    A_s=U_s\Lambda U^{T}_s
\end{equation}
From (\ref{eq64}), we let
\begin{equation}\label{eq40}
    U=D^{-1/2}U_s,\quad U^{-1}=U^{T}_sD^{1/2}
\end{equation}
and we obtain (\ref{eq67}).
\end{proof}}
Note that since the matrices $U_s$ and $D^{1/2}$ in (\ref{eq40}) are
real-valued, so are eigenvectors of the uniform combination matrix,
$\{r_k,s_k\}$. Furthermore, from (\ref{eq40}), we can express
$\{r_k,s_k\}$ in terms of the eigenvectors of $A_s$ defined in
(\ref{eq64}). Let $r^s_{k}$ denote the $k$th eigenvector of $A_s$
and let $r^s_{k,l}$ denote the $l$th entry of $r^s_{k}$. Likewise,
let $\{r_{k,l}, s_{k,l}\}$ denote the $l$th entries of
$\{r_k,s_k\}$. Then, we have
\begin{equation}\label{eq93}
    r_{k,l}=\frac{r^s_{k,l}}{\sqrt{n_l}},\quad
    s_{k,l}=\sqrt{n_l}\cdot r^s_{k,l}
\end{equation}
For the Erdos-Renyi model, since nodes have on average the same
expected degree given by (\ref{eq86}), i.e., $n_k\approx
\bar{n}_k=\bar{\eta}$, then the right eigenvectors $\{r_k\}$ of the
uniform combination matrix defined by (\ref{eq21}) are approximately
orthogonal in view of
\begin{equation}\label{eq59}
\begin{aligned}
    \left|r^T_lr_k\right|=\left|\sum_{m=1}^N\frac{r^s_{l,m}r^s_{k,m}}{n_m}\right|
    \approx \frac{1}{\bar{\eta}}\left|\sum_{m=1}^Nr^s_{l,m}r^s_{k,m}\right|
    =\frac{1}{\bar{\eta}}\delta_{kl}
\end{aligned}
\end{equation}
Approximation (\ref{eq59}) is particularly good when $N$ is large
since most nodes have degree similar to $\bar{\eta}$. Even though
this approximation is not generally valid for the scale-free model,
simulations further ahead indicate that the approximation still
leads to good match between theory and practice.
\begin{rem}{\rm
We note that for networks with random degree distributions, such as
the Erdos-Renyi and scale-free networks of Sec. IV, the matrix $A$
is generally a random matrix. In the sequel, we shall derive
expressions for the convergence rate and network MSD for}
realizations {\rm of the network
--- see expressions (\ref{eq43}) and (\ref{eq42}) further ahead. To
evaluate the} expected {\rm convergence rate and network MSD over a
probability distribution for the degrees (such as (\ref{eq33}) or
(\ref{eq19})), we will need to average expressions (\ref{eq43}) and
(\ref{eq42}) over the degree distribution.} \vspace{-1cm}
\begin{flushright}
    $\blacksquare$
\end{flushright}
\end{rem}

For the covariance matrix $R_u$, we let $z_m$ ($m=1,\ldots,M$)
denote the eigenvector of $R_u$ that is associated with the
eigenvalue $\lambda_m(R_u)$. Then, the eigen-decomposition of $R_u$
is given by:
\begin{equation}\label{eq66}\boxed{
    R_u=\sum_{m=1}^M\lambda_m(R_u)\cdot z_mz_m^*}
\end{equation}
where the $\{z_m\}$ are orthonormal, i.e., $z^*_nz_m=\delta_{mn}$,
and the $\{\lambda_m(R_u)\}$ are again arranged in decreasing order
with $\lambda_1(R_u)\geq\lambda_2(R_u)\geq\cdots
\geq\lambda_M(R_u)>0$. In the sequel, for any vector $x$, we use the
notation $x_{k:l}$ to denote a sub-vector of $x$ formed from the
$k$th up to the $l$th entries of $x$. Also, we let $N_I$ denote the
number of informed nodes in the network. Without loss of generality,
we label the network nodes such that the first $N_I$ nodes are
informed, i.e., $\mathcal{N}_I=\{1,2,\ldots,N_I\}$. The next result
establishes a useful approximation for the eigen-structure of the
matrix $\mathcal{B}$ defined in (\ref{eq9}); it shows how the
eigenvectors and eigenvalues of $\mathcal{B}$ can be constructed
from the eigenvalues and eigenvectors for $\{A^T,R_u\}$ given by
(\ref{eq60}) and (\ref{eq66}).
\begin{lem}[Eigen-structure of $\mathcal{B}$]
For a symmetric ATC network (\ref{eq49}) with at least one informed
node, the matrix
$\mathcal{B}=\mathcal{A}^T(I-\mathcal{M}\mathcal{R})$ has
approximate right and left eigenvector pairs
$\{r^b_{k,m},s^{b}_{k,m}\}$ given by:
\begin{align}\label{eq11}
    r^b_{k,m}&\approx r_k\otimes z_m,\;\;\; k=1,\ldots,N;\;\;m=1,\ldots,M\\
    s^{b*}_{k,m}&\approx
    \frac{\lambda_k(A)}{\lambda_{k,m}(\mathcal{B})}\cdot
    \begin{bmatrix}\left(1-\mu\lambda_m(R_u)\right)\cdot s^*_{k,1:N_I}\otimes z^*_m
    & s^*_{k,N_I+1:N}\otimes z^*_m\end{bmatrix} \label{eq52}
\end{align}
where $\lambda_{k,m}(\mathcal{B})$ denotes the eigenvalue of the
eigenvector pair $\{r^b_{k,m},s^{b}_{k,m}\}$ and is approximated by:
\begin{equation}\label{eq56}
    \lambda_{k,m}(\mathcal{B})\approx\lambda_k(A)\cdot
    \left[1-\mu\lambda_m(R_u)\cdot s^*_{k,1:N_I}r_{k,1:N_I}\right]
\end{equation}
\end{lem}
{
\begin{proof}
We first note from (\ref{eq96}) and (\ref{eq60}) that the matrix
$\mathcal{A}^T$ can be written as
\begin{equation}\label{eq61}
    \mathcal{A}^T=\sum_{l=1}^N\lambda_l(A)(r_l\otimes
    I_M)(s_l^*\otimes I_M)
\end{equation}
Then, the matrix $\mathcal{B}$ in (\ref{eq9}) becomes
\begin{equation}\label{eq62}
\begin{aligned}
    \mathcal{B}&=\sum_{l=1}^N\lambda_l(A)(r_l\otimes I_M)(s^*_l\otimes
    I_M)\left(I_{NM}-\begin{bmatrix}\mu I_{N_IM} & \\
    & 0_{(N-N_I)M} \end{bmatrix}(I_N\otimes R_u)\right)\\
    &=\sum_{l=1}^N\lambda_l(A)(r_l\otimes I_M)
    \begin{bmatrix}s^*_{l,1:N_I}\otimes (I_M-\mu R_u) &
    s^*_{l,N_I+1:N}\otimes I_M\end{bmatrix}
\end{aligned}
\end{equation}
Multiplying $\mathcal{B}$ by the $r^b_{k,m}$ defined in (\ref{eq11})
from the right, we obtain
\begin{align}
    \mathcal{B}\cdot r^b_{k,m}&=\sum_{l=1}^N\lambda_l(A)\cdot(r_l\otimes I_M)
    \left[ s^*_{l,1:N_I}r_{k,1:N_I} \otimes (1-\mu R_u)z_m +
    s^*_{l,N_I+1:N}r_{k,N_I+1:N} \otimes z_m \right]\notag \\
    &=\sum_{l=1}^N\lambda_l(A)\cdot\left[\left(1-\mu
    \lambda_m(R_u)\right)
    s^*_{l,1:N_I}r_{k,1:N_I}+s^*_{l,N_I+1:N}r_{k,N_I+1:N}\right]
    (r_l\otimes I_M)(1\otimes z_m)\notag \\
    &=\sum_{l=1}^N\lambda_l(A)\cdot\left[s^*_{l}r_{k} -\mu
    \lambda_m(R_u)\cdot
    s^*_{l,1:N_I}r_{k,1:N_I}\right]\cdot(r_l\otimes z_m)\notag \\
    &\begin{aligned}
    =\lambda_k(A)\cdot
    \left[1 -\mu \lambda_m(R_u)\cdot
    s^*_{k,1:N_I}r_{k,1:N_I}\right]\cdot(r_k\otimes z_m)\\
    \phantom{==}-\mu \lambda_m(R_u)\sum_{l\neq k}\lambda_l(A)\cdot
    s^*_{l,1:N_I}r_{k,1:N_I}\cdot(r_l\otimes z_m)
    \end{aligned}\label{eq51}
\end{align}
where we used that $s^*_{l}r_{k}= \delta_{kl}$. For sufficiently
small step-sizes, we can ignore the second term in the last equation
of (\ref{eq51}) and write:
\begin{equation}\label{eq97}
\begin{aligned}
    \mathcal{B}\cdot r^b_{k,m}&\approx \lambda_k(A)\cdot
    \left[1 -\mu \lambda_m(R_u)\cdot
    s^*_{k,1:N_I}r_{k,1:N_I}\right]\cdot(r_k\otimes z_m)\\
    &=\lambda_{k,m}(\mathcal{B})\cdot r^b_{k,m}
\end{aligned}
\end{equation}
Note that approximation (\ref{eq97}) is particularly good for the
uniform combination matrix in (\ref{eq21}) since, from (\ref{eq93})
and by the Cauchy-Schwarz inequality, we have
\begin{equation}
\begin{aligned}
    |s^*_{l,1:N_I}r_{k,1:N_I}|=\left|\sum_{m=1}^{N_I}r^s_{l,m}r^s_{k,m}\right|
    \leq \left|\sum_{m=1}^{N_I}(r^s_{k,m})^2\right|
    = |s^*_{k,1:N_I}r_{k,1:N_I}|
\end{aligned}
\end{equation}
Following similar arguments, we can verify that
\begin{equation}
\begin{aligned}
    s^{b*}_{k,m}\cdot \mathcal{B}&=\frac{\lambda_k(A)}{\lambda_{k,m}(\mathcal{B})}\cdot
    \sum_{l=1}^N\lambda_l(A)\cdot \left[s^*_{l}r_{k} -\mu \lambda_m(R_u)\cdot
    s^*_{l,1:N_I}r_{k,1:N_I}\right]\\
    &\phantom{==}\times
    (1\otimes z^*_m)\begin{bmatrix}s^*_{l,1:N_I}\otimes (I_M-\mu R_u) &
    s^*_{l,N_I+1:N}\otimes I_M\end{bmatrix}\\
    &\approx\frac{\lambda_k(A)}{\lambda_{k,m}(\mathcal{B})}\cdot
    \lambda_{k,m}(\mathcal{B})\cdot
    \begin{bmatrix}\left(1-\mu\lambda_m(R_u)\right)s^*_{k,1:N_I}\otimes z^*_m &
    s^*_{k,N_I+1:N}\otimes z^*_m\end{bmatrix}\\
    &=\lambda_{k,m}(\mathcal{B})\cdot s^{b*}_{k,m}
\end{aligned}
\end{equation}
\end{proof}}
\noindent Now, we argue that the approximate eigenvalues of
$\mathcal{B}$ in (\ref{eq56}) have magnitude less than one, i.e.,
$|\lambda_{k,m}(\mathcal{B})|<1$ for all $k$ and $m$. Note that,
since $|\lambda_k(A)|<1$ for $k>1$ and for sufficiently small
step-sizes, we have $|\lambda_{k,m}(\mathcal{B})|\approx
|\lambda_k(A)|<1$ for $k>1$. For $k=1$, $\lambda_1(A)=1$. However,
since the eigenvectors $\{r_1,s_1\}$ have all positive entries, as
we remarked before, we have $0<s^*_{1,1:N_I}r_{1,1:N_I}\leq
s^*_{1}r_{1}=1$. In addition, from Assumption 1 that $0<\mu
\rho(R_u)<1$ and $\lambda_m(R_u)>0$ for all $m$, we know that
\begin{equation}\label{eq91}
\begin{aligned}
    0< 1-\mu\rho(R_u)
    \leq 1-\mu\lambda_m(R_u)\cdot s^*_{1,1:N_I}r_{1,1:N_I}
    < 1
\end{aligned}
\end{equation}
and we conclude that $|\lambda_{1,m}(\mathcal{B})|<1$ for all $m$.
For the uninform combination matrix defined in (\ref{eq21}), since
all eigenvectors and eigenvalues of $A$ are real-valued, we further
have that the $\{\lambda_{k,m}(\mathcal{B})\}$ are real.

\subsection{Simplifying the MSD Expression (\ref{eq31})}
Using the result of Lemma 4, we find that the eigen-decomposition
for the matrix $\mathcal{B}^j$ has the approximate form:
\begin{equation}\label{eq87}\boxed{
    \mathcal{B}^j\approx\sum_{k=1}^N\sum_{m=1}^M\lambda^j_{k,m}(\mathcal{B})
    \cdot r^b_{k,m}s^{b*}_{k,m}}
\end{equation}
we can rewrite the network MSD (\ref{eq31}) in the form:
\begin{equation}\label{eq20}
\begin{aligned}
    \text{MSD}&\approx\frac{1}{N}
    \sum_{j=0}^\infty\sum_{k,l=1}^N\sum_{m,n=1}^M
    \text{Tr}\left[\lambda^j_{k,m}(\mathcal{B})\lambda^{*j}_{l,n}(\mathcal{B})
    \cdot r^b_{k,m}s^{b*}_{k,m}
    \mathcal{Y}s^{b}_{l,n}r^{b*}_{l,n}\right]\\
    &=\sum_{k,l=1}^N\sum_{m,n=1}^M
    \frac{\left(r^{b*}_{l,n}r^b_{k,m}\right)\cdot
    \left(s^{b*}_{k,m} \mathcal{Y}s^{b}_{l,n}\right)}
    {N\cdot \left[1-\lambda_{k,m}(\mathcal{B})\lambda^*_{l,n}(\mathcal{B})\right]}
\end{aligned}
\end{equation}
Moreover, from (\ref{eq11}) and assumption (\ref{eq10}), since
\begin{equation}
\begin{aligned}
    r^{b*}_{l,n}r^b_{k,m}&= \left(r^*_lr_k\right)\otimes \left(z^*_nz_m\right)\\
    &\approx\|r_k\|^2\cdot\delta_{kl}\cdot\delta_{mn}
\end{aligned}
\end{equation}
expression (\ref{eq20}) simplifies to:
\begin{equation}\label{eq26}
\begin{aligned}
    \text{MSD}\approx\sum_{k=1}^N\sum_{m=1}^M
    \frac{\|r_k\|^2\cdot s^{b*}_{k,m} \mathcal{Y}s^{b}_{k,m}}
    {N\cdot [1-|\lambda_{k,m}(\mathcal{B})|^2]}
\end{aligned}
\end{equation}
Expression (\ref{eq26}) can be simplified further once we evaluate
the term in the numerator. We start by expressing the matrix
$\mathcal{Y}$ from (\ref{eq25}) as:
\begin{equation}
    \mathcal{Y} = \mathcal{Z}\Omega^{-1}\mathcal{Z}^*
\end{equation}
where
\begin{align}\label{eq53}
    \mathcal{Z} &= \mathcal{A}^T \mathcal{M}\mathcal{R}\\
    \Omega &= \text{diag}\{\sigma^{-2}_{v,1}R_u,\ldots,\sigma^{-2}_{v,N}R_u\}
    =\Sigma_v^{-1}\otimes R_u
\end{align}
with $\Sigma_v\triangleq
\text{diag}\{\sigma^2_{v,1},\ldots,\sigma^2_{v,N}\}$. Then, we get
\begin{equation}\label{eq55}
    s^{b*}_{k,m} \mathcal{Y}s^{b}_{k,m}= \|s^{b*}_{k,m}\mathcal{Z}\Omega^{-1/2}\|^2
\end{equation}
Note from (\ref{eq61}) and (\ref{eq62}) that the matrix
$\mathcal{Z}$ in (\ref{eq53}) can be written as:
\begin{equation}\label{eq54}
\begin{aligned}
    \mathcal{Z}=\mathcal{A}^T-\mathcal{B}
    =\sum_{l=1}^N\lambda_l(A)(r_l\otimes I_M)
    \begin{bmatrix}s^*_{l,1:N_I}\otimes \mu R_u &
    s^*_{l,N_I+1:N}\otimes 0_{M}\end{bmatrix}
\end{aligned}
\end{equation}
We then obtain from (\ref{eq52}), (\ref{eq53}), and (\ref{eq54})
that:
\begin{equation}
\begin{aligned}
    s^{b*}_{k,m}\mathcal{Z}\Omega^{-1/2}&=
    \sum_{l=1}^N\lambda_l(A)\cdot s^{b*}_{k,m}(r_l\otimes I_M)\cdot
    \begin{bmatrix}s^*_{l,1:N_I}\otimes \mu R_u &
    s^*_{l,N_I+1:N}\otimes 0_{M}\end{bmatrix}\Omega^{-1/2}\\
    &\approx
    \frac{\lambda_k(A)}{\lambda_{k,m}(\mathcal{B})}\cdot\lambda_{k,m}(\mathcal{B})
    \cdot(1\otimes z^*_m)\begin{bmatrix}s^*_{k,1:N_I}\Sigma^{1/2}_{v,1:N_I}
    \otimes \mu R^{1/2}_u & 0_{1\times (N-N_I)M}\end{bmatrix}\\
    &=
    \lambda_k(A)\cdot\begin{bmatrix}s^*_{k,1:N_I}\Sigma^{1/2}_{v,1:N_I}
    \otimes \mu \lambda^{1/2}_m(R_u)z^*_m &
    0_{1\times (N-N_I)M}\end{bmatrix}
\end{aligned}
\end{equation}
Therefore, the term $s_{k,m}^{b*}\mathcal{Y}s^b_{k,m}$ in
(\ref{eq55}) becomes
\begin{equation}\label{eq57}
\begin{aligned}
    s_{k,m}^{b*}\mathcal{Y}s^b_{k,m} &=
    \left(s^{b*}_{k,m}\mathcal{Z}\Omega^{-1/2}\right)
    \left(s^{b*}_{k,m}\mathcal{Z}\Omega^{-1/2}\right)^*\\
    &\approx
    \mu^2\lambda_m(R_u)|\lambda_k(A)|^2\cdot
    s^*_{k,1:N_I}\Sigma_{v,1:N_I}s_{k,1:N_I}
\end{aligned}
\end{equation}
where we used that $z^*_mz_m=1$. Then, substituting (\ref{eq56}) and
(\ref{eq57}) into (\ref{eq26}), we arrive at the following
expression for the network MSD in terms of the eigenvalues and
eigenvectors of $A^T$ and the eigenvalues of $R_u$.
\begin{thm}[Network MSD]
The network MSD of the ATC strategy (\ref{eq49}) can be
approximately expressed as
\begin{equation}\label{eq22}\boxed{
    {\rm{MSD}}\approx\sum_{k=1}^N\sum_{m=1}^M
    \frac{\mu^2\lambda_m(R_u)|\lambda_k(A)|^2\cdot\|r_k\|^2\cdot
    s^*_{k,1:N_I}\Sigma_{v,1:N_I}s_{k,1:N_I}}
    {N\left[1-|\lambda_{k}(A)|^2\cdot \left|1-
    \mu\lambda_m(R_u)\cdot s^*_{k,1:N_I}r_{k,1:N_I}\right|^2\right]}}
\end{equation}\vspace{-1cm}
\begin{flushright}
    $\blacksquare$
\end{flushright}
\end{thm}
Since the matrix $A$ has a single eigenvalue at $\lambda_1(A)=1$,
and its value is greater than the remaining eigenvalues, we can
decompose the MSD in (\ref{eq22}) into two components. The first
component is determined by $\lambda_1(A)$, i.e., $k=1$ in
(\ref{eq22}), and is denoted by $\text{MSD}_{k=1}$. The second
component is due to the contribution from the remaining eigenvalues
of $A$, i.e., $k>1$ in (\ref{eq22}), and is denoted by
$\text{MSD}_{k>1}$. Since $\lambda_1(A)=1$, and for sufficiently
small step-sizes, we introduce the approximation for the denominator
in (\ref{eq22}):
\begin{equation}\label{eq98}
    |\lambda_{1}(A)|^2\cdot \left|1-
    \mu\lambda_m(R_u)\cdot s^*_{1,1:N_I}r_{1,1:N_I}\right|^2
    \approx 1-2\mu\lambda_m(R_u)\cdot s^T_{1,1:N_I}r_{1,1:N_I}
\end{equation}
Then, the term $\text{MSD}_{k=1}$ becomes
\begin{equation}\label{eq23}\boxed{
\begin{aligned}
    \text{MSD}_{k=1}&\approx\frac{M\mu\|r_1\|^2}{2N}\cdot
    \frac{\sum_{l=1}^{N_I}\sigma^2_{v,l}s^2_{1,l}}
    {\sum_{l=1}^{N_I}r_{1,l}s_{1,l}}
\end{aligned}}
\end{equation}
For the second part, $\text{MSD}_{k>1}$, since $|\lambda_k(A)|<1$
for $k>1$, and for sufficiently small step-sizes, the denominator in
(\ref{eq22}) can be approximated by:
\begin{equation}\label{eq99}
\begin{aligned}
    1-|\lambda_k(A)|^2\cdot\left|1-
    \mu\lambda_m(R_u)\cdot s^*_{k,1:N_I}r_{k,1:N_I}\right|^2
    \approx 1-|\lambda_k(A)|^2
\end{aligned}
\end{equation}
Comparing to (\ref{eq98}), we further ignore the term
$2\mu\lambda_m(R_u)|\lambda_k(A)|^2\cdot s^*_{k,1:N_I}r_{k,1:N_I}$
in (\ref{eq99}) since this term is generally much less than
$1-|\lambda_k(A)|^2$, especially for well-connected networks, i.e.,
high value of $\bar{\eta}$ (see (\ref{eq72}) further ahead). Then,
$\text{MSD}_{k>1}$ becomes
\begin{equation}\label{eq63}\boxed{
\begin{aligned}
    \text{MSD}_{k>1}\approx
    \frac{\mu^2\text{Tr}(R_u)}{N}\sum_{k=2}^N\left[
    \frac{|\lambda_k(A)|^2\cdot\|r_k\|^2}{1-|\lambda_{k}(A)|^2}\cdot
    \sum_{l=1}^{N_I}\sigma^2_{v,l}|s_{k,l}|^2\right]
\end{aligned}}
\end{equation}
As shown by (\ref{eq22}), (\ref{eq23}), and (\ref{eq63}), the
network MSD depends strongly on the eigenvalues and eigenvectors of
the combination matrix $A$. In the next section, we examine more
closely the eigen-structure of the uniform combination matrix $A$
from (\ref{eq21}). In a subsequent section, we employ the results to
assess how the MSD varies with the proportion of informed nodes
--- see expressions (\ref{eq58}) and (\ref{eq32}) further ahead.

\subsection{MSD Expression for the Uniform Combination Matrix from (\ref{eq21})}
\subsubsection*{C.1) Eigenvalues of $A$}
\begin{figure}
\centering
\includegraphics[width=34em]{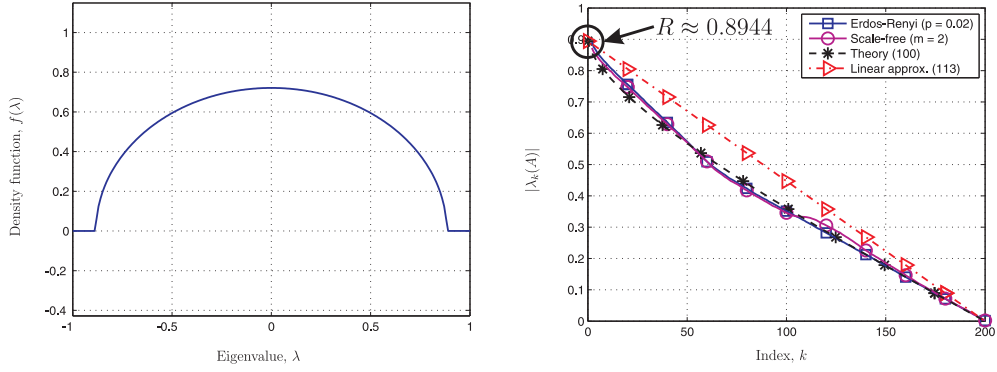}
\caption{Density function (left) for the eigenvalues of $A$ as given
by (\ref{eq68}) for $N\rightarrow \infty$, and averaged eigenvalues
(right) of the combination matrix $A$ defined by (\ref{eq21}) over
$30$ experiments with $\eta=5$. The dashed line on the right
represents theory from (\ref{eq74}) and the dash-dot line represents
linear approximation given further ahead by (\ref{eq37}).}
\label{Fig_2}
\end{figure}

We start by examining the eigenvalues of the uniform combination
matrix $A$ from (\ref{eq21}). We define the Laplacian matrix, $L$,
of a network graph as:
\begin{equation}
    L \triangleq D-C
\end{equation}
in terms of the $D$ and $C$ from (\ref{eq65}). Then, the normalized
Laplacian matrix is defined as \cite{Chung03}:
\begin{equation}
\begin{aligned}
    \mathcal{L}\triangleq D^{-1/2}LD^{-1/2}
    =I-A_s
\end{aligned}
\end{equation}
where $A_s$ is the same matrix defined earlier in (\ref{eq64}). From
Lemma \ref{lem_2}, we know that the matrices $A$ and $A_s$ have the
same eigenvalues and we conclude that
\begin{equation}
    \lambda_k(\mathcal{L})=1-\lambda_k(A)
\end{equation}
In other words, the spectrum of $A$ is related to the spectrum of
the normalized Laplacian matrix. There are useful results in the
literature on the spectral properties of the Laplacian matrices for
random graphs \cite{Farkas01,Goh01,Chung03,Dorogovtsev04}, such as
the graphs corresponding to the Erdos-Renyi and scale free models of
Sec. IV. We shall use these results to infer properties about the
spectral distribution of the corresponding combination matrices $A$
that are defined by (\ref{eq21}). In particular, reference
\cite{Chung03} gives an expression for the eigenvalue distribution
of $\mathcal{L}$ for certain random graphs; this expression can be
used to infer the eigenvalue distribution of $A$, as we now verify.
First we note from (\ref{eq50}) that one is an eigenvalue of $A$,
i.e., $\rho(A)=\lambda_1(A)=1$. In the following, we use the results
of \cite{Chung03} to characterize the remaining eigenvalues (namely,
$\lambda_k(A)$ for $k>1$) of uniform combination matrix.
\begin{thm}[Eigenvalue distribution of $A$]
Let $\bar{n}_k$ denote the average degree of node $k$ in a random
graph. Let
\begin{equation}\label{eq27}
    \bar{\eta} \triangleq \frac{1}{N} \sum_{k=1}^N \bar{n}_k
\end{equation}
denote the average degree of the graph. Then, for random graphs with
expected degrees satisfying
\begin{equation}\label{eq71}
    \bar{n}_{\min}\triangleq\min_{1\leq k\leq N}\{\bar{n}_k\}
    \gg \sqrt{\bar{\eta}}
\end{equation}
the density function, $f(\lambda)$, of the eigenvalues of $A$
converges in probability, as $N\rightarrow \infty$, to the
semicircle law (see Fig. \ref{Fig_2}), i.e.,
\begin{equation}\label{eq68}
    f(\lambda) =
    \begin{cases}
    \frac{2}{\pi R}\sqrt{1-\left(\frac{\lambda}{R}\right)^2},
    &\text{if $\lambda \in[-R,R]$}\\
    0, &\text{otherwise}
    \end{cases}
\end{equation}
where
\begin{equation}\label{eq94}
    R = \frac{2}{\sqrt{\bar{\eta}}}
\end{equation}
Moreover, if $\bar{n}_{\min}\gg \sqrt{\bar{\eta}}\log^3(N)$, the
second largest eigenvalue of $A$ converges almost surely to
\begin{equation}\label{eq72}
    |\lambda_2(A)|=R
\end{equation}
\end{thm}
{
\begin{proof}
See Thms. 5 and 6 in \cite{Chung03}.
\end{proof}}
Simulations further ahead (see Fig. \ref{Fig_2}) show that
expressions (\ref{eq68}) and (\ref{eq72}) provide accurate
approximations for the Erdos-Renyi and scale-free network models
described in Section IV. In addition, for ergodic distributions, the
value of $\bar{\eta}$ in (\ref{eq27}) will be close to its
realization $\eta$ for large $N$, where $\eta$ is defined as
\begin{equation}\label{eq46}\boxed{
    \eta\triangleq\frac{1}{N}\sum_{k=1}^Nn_k}
\end{equation}
In the following, we determine an expression for $|\lambda_k(A)|$ by
using (\ref{eq68}). To do so, we let $k$ denote the number of
eigenvalues of $A$ that are greater than some value $y$ in magnitude
for $0\leq y\leq R$. Then, the value of $k$ is given by:
\begin{equation}\label{eq69}
\begin{aligned}
    k &= N\cdot \left[1-\int_{-y}^yf(\lambda)d\lambda\right]\\
    &\triangleq N\cdot g(y)
\end{aligned}
\end{equation}
where we denote the expression inside the brackets by $g(y)$. Note
that the integral $\int_{-y}^yf(\lambda)d\lambda$ in (\ref{eq69})
computes the proportion of eigenvalues of $A$ within the region
$[-y,y]$. Then, the $k$th eigenvalue of $A$ can be approximated by
evaluating the value of $y$ in (\ref{eq69}), i.e.,
\begin{equation}\label{eq74}
    |\lambda_k(A)| \approx g^{-1}\left(\frac{k}{N}\right)
\end{equation}
From (\ref{eq68}) and using the change of variables
$\lambda/R=\sin\theta$, we obtain that $g(y)$ in (\ref{eq69}) has
the form:
\begin{equation}\label{eq73}
\begin{aligned}
    g(y) = 1-\frac{2}{\pi}\sin^{-1}\left(\frac{y}{R}\right)
    -\frac{2}{\pi}\frac{y}{R}\sqrt{1-\left(\frac{y}{R}\right)^2}
\end{aligned}
\end{equation}
In Fig. \ref{Fig_2}, we show the averaged distribution of
$|\lambda_k(A)|$ for Erdos-Renyi and scale-free models over 30
experiments. We observe that for both network models, the
theoretical results in (\ref{eq72}) and (\ref{eq74}) match well with
simulations.

\subsubsection*{C.2) MSD Expression for $k=1$}
From (\ref{eq23}), MSD$_{k=1}$ depends on the eigenvectors
$\{r_{1},s_1\}$. For the uniform combination matrix $A$ in
(\ref{eq21}), it can be verified that the right eigenvector for
$A_s$ defined in (\ref{eq64}) corresponding to the eigenvalue one
has the following form:
\begin{equation}\label{eq48}
    r^s_1 = \frac{1}{\sqrt{N\eta}}
    \text{col}\{\sqrt{n_1},\ldots,\sqrt{n_N}\}
\end{equation}
Then, from (\ref{eq93}) and (\ref{eq48}), expression (\ref{eq23})
becomes
\begin{equation}\label{eq58}\boxed{
\begin{aligned}
    \text{MSD}_{k=1}\approx
    \frac{M\mu}{2N}\cdot
    \frac{\sum_{l=1}^{N_I}\sigma^2_{v,l}n^2_l}
    {\eta \sum_{l=1}^{N_I}n_l}
\end{aligned}}\quad\text{(using uniform combination matrix (\ref{eq21}))}
\end{equation}
Expression (\ref{eq58}) reveals several interesting properties.
First, we observe that the term $\text{MSD}_{k=1}$ does not depend
on the matrix $R_u$, which is also a property of the MSD expression
for stand-alone adaptive filters \cite{Sayed08}. Second, expression
(\ref{eq58}) is inversely proportional to the degree of the network
realization, $\eta$. That is, when the network is more connected
(e.g., higher values of $p$ and $m$ in the Erdos-Renyi and
scale-free models), the network will have lower MSD$_{k=1}$. Third,
expression (\ref{eq58}) depends on the distribution of
\emph{informed} nodes through its dependence on the degree and noise
profile of the informed nodes. \emph{If the number of informed nodes
increases by one, the value of} MSD$_{k=1}$ \emph{may increase or
decrease} (i.e., it does not necessarily decrease). This can be seen
as follows. From (\ref{eq58}) we see that $\text{MSD}_{k=1}$ will
decrease (and, hence, improve) only if
\begin{equation}\label{eq79}
    \frac{\sum_{l=1}^{N_I}\sigma^2_{v,l}n^2_l+\sigma^2_{v,N_I+1}n^2_{N_I+1}}
    {\sum_{l=1}^{N_I}n_l+n_{N_I+1}}<
    \frac{\sum_{l=1}^{N_I}\sigma^2_{v,k}n^2_l}
    {\sum_{l=1}^{N_I}n_l}
\end{equation}
or, if the degree of the added node satisfies:
\begin{equation}\label{eq29}
    \sigma^2_{v,N_I+1}n_{N_I+1}<\frac{\sum_{l=1}^{N_I}\sigma^2_{v,k}n^2_l}
    {\sum_{l=1}^{N_I}n_l}
\end{equation}

\subsubsection*{C.3) MSD Expression for $k>1$}
\begin{figure}
\centering
\includegraphics[width=34em]{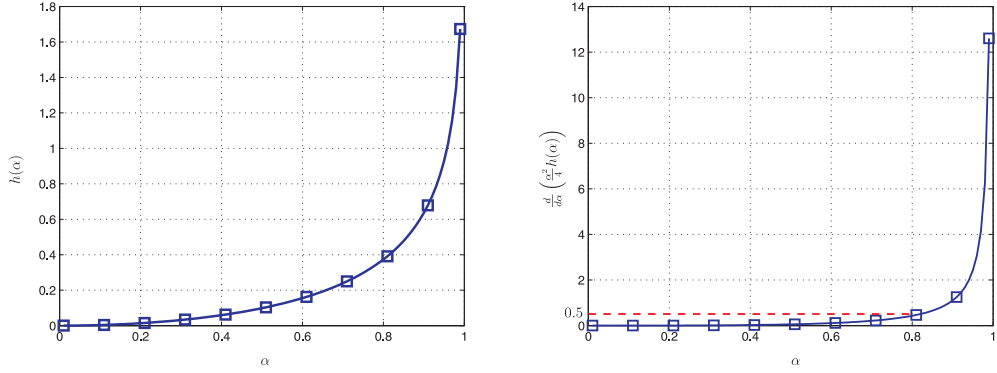}
\caption{The function $h(\alpha)$ (left) from (\ref{eq23}) and the
derivative of $\alpha^2h(\alpha)/4$ with respect to $\alpha$
(right).} \label{Fig_3}
\end{figure}

For MSD$_{k>1}$, we apply relation (\ref{eq93}) and approximation
(\ref{eq59}). Then, expression (\ref{eq63}) can be approximated by:
\begin{equation}\label{eq34}
\begin{aligned}
    \text{MSD}_{k>1}=\frac{\mu^2\text{Tr}(R_u)}{N\eta}\sum_{k=2}^N\left[
    \frac{\lambda^2_k(A)}{1-\lambda^2_{k}(A)}\cdot
    \left(\sum_{l=1}^{N_I}\sigma^2_{v,l}n_l\cdot(r^s_{k,l})^2\right)\right]
\end{aligned}
\end{equation}
where we replaced $\bar{\eta}$ by $\eta$ for large $N$. Expression
(\ref{eq34}) requires knowledge of the eigenvectors $\{r^s_k\}$ of
$A_s$ in (\ref{eq64}). Note that for $k=1$ and from (\ref{eq48}), we
have
\begin{equation}
    (r^s_{1,l})^2=\frac{n_l}{N\eta}\approx \frac{1}{N}
\end{equation}
since the nodes have similar degree in the Erdos-Renyi model. We are
therefore motivated to introduce the following approximation:
\begin{equation}\label{eq45}
    (r^s_{k,l})^2\approx\frac{1}{N}
\end{equation}
for all $k$. Observe that expression (\ref{eq45}) is independent of
$k$, and we find that expression (\ref{eq34}) simplifies to:
\begin{equation}\label{eq36}
\begin{aligned}
    \text{MSD}_{k>1}\approx
    \frac{\mu^2\text{Tr}(R_u)}{N\eta}\cdot
    \left(\sum_{l=1}^{N_I}\sigma^2_{v,l}n_l\right)\cdot\frac{1}{N}
    \sum_{k=2}^N \frac{\lambda^2_k(A)}{1-\lambda^2_{k}(A)}
\end{aligned}
\end{equation}
Furthermore, from (\ref{eq74}), we can approximate the summation
over $k$ in (\ref{eq36}) by the following integral:
\begin{equation}\label{eq75}
\begin{aligned}
    \frac{1}{N}\sum_{k=2}^N\frac{\lambda^2_k(A)}{1-\lambda^2_k(A)}
    &\approx
    \int_0^{1}\frac{\left[g^{-1}(x)\right]^2}{1-\left[g^{-1}(x)\right]^2}dx
\end{aligned}
\end{equation}
where we replaced $k/N$ by $x$. However, evaluating the integral in
(\ref{eq75}) is generally intractable. We observe though from the
right plot in Fig. \ref{Fig_2} that $|\lambda_k(A)|$ (and also
$g^{-1}(k/N)$) decreases in a rather linear fashion for $k>1$. Note
that the function $g(y)$ in (\ref{eq73}) has values $1$ at $y=0$ and
$0$ at $y=R\approx 2/\sqrt{\eta}$. We therefore approximate $g(y)$
by the linear function
\begin{equation}\label{eq37}
    g(y) \approx 1-\frac{\sqrt{\eta}}{2}y
\end{equation}
Then,
\begin{equation}\label{eq37}
    g^{-1}(x)\approx\frac{2}{\sqrt{\eta}}(1-x)
\end{equation}
and expression (\ref{eq75}) becomes
\begin{equation}\label{eq24}
\begin{aligned}
    \frac{1}{N}\sum_{k=2}^N\frac{\lambda^2_k(A)}{1-\lambda^2_k(A)}
    &\approx
    \int_0^{1}\frac{4/\eta\cdot (1-x)^2}{1-4/\eta\cdot (1-x)^2}dx\\
    &=h\left(\frac{2}{\sqrt{\eta}}\right)
\end{aligned}
\end{equation}
where the function $h(\alpha)$ is defined as
\begin{equation}\label{eq38}
    h(\alpha)\triangleq\left[\frac{1}{2\alpha}
    \log\left(\frac{1+\alpha}{1-\alpha}\right)-1\right]
\end{equation}
Substituting expression (\ref{eq24}) into (\ref{eq36}), we find that
the MSD contributed by the remaining terms ($k>1$) has the following
form:
\begin{equation}\label{eq32}\boxed{
\begin{aligned}
    \text{MSD}_{k>1}\approx
    \frac{\mu^2\text{Tr}(R_u)}{N\eta}\cdot
    \left(\sum_{l=1}^{N_I}\sigma^2_{v,l}n_l\right)\cdot
    h\left(\frac{2}{\sqrt{\eta}}\right)
\end{aligned}}\quad \text{(using uniform combination matrix (\ref{eq21}))}
\end{equation}
Note that, in contrast to MSD$_{k=1}$ in (\ref{eq58}), MSD$_{k>1}$
\emph{in (\ref{eq32}) always increases when the number of informed
nodes increases.} Moreover, the function $h(\alpha)$, shown in Fig.
\ref{Fig_3}, has the following property.

\begin{lem}\label{lem3}
The function $h(\alpha)$ defined in (\ref{eq38}) is strictly
increasing and convex in $\alpha\in(0,1)$.
\end{lem}
{
\begin{proof}
From (\ref{eq24}), we note that $h(\alpha)$ can be written in the
integral form:
\begin{equation}\label{eq39}
    h(\alpha)=\int_0^1 \frac{\alpha^2x^2}{1-\alpha^2x^2}dx
\end{equation}
Taking the derivative of $h(\alpha)$ in (\ref{eq39}) with respect to
$\alpha$, we obtain:
\begin{equation}
    \frac{dh(\alpha)}{d\alpha}=\int_0^1\frac{2\alpha
    x^2}{(1-\alpha^2x^2)^2}dx>0
\end{equation}
for $\alpha\in(0,1)$. To show convexity, we take the second
derivative of $h(\alpha)$ for $\alpha\in(0,1)$ and find that
\begin{equation}
    \frac{d^2h(\alpha)}{d\alpha^2}=\int_0^1\frac{2x^2+6\alpha^2
    x^4}{(1-\alpha^2x^2)^3}dx>0
\end{equation}
\end{proof}}
The result of Lemma \ref{lem3} implies that when $\eta$ (or, $p$ or
$m$) increases, $\text{MSD}_{k>1}$ in (\ref{eq32}) decreases. That
is, in a manner similar to $\text{MSD}_{k=1}$ in (\ref{eq58}), the
value of $\text{MSD}_{k>1}$ is lower if the network is more
connected. In addition, we observe that when $\eta$ is too low (or,
$\alpha$ is too large in Fig. \ref{Fig_3}), the value of
$h(2/\sqrt{\eta})$ will increase rapidly and so does the value of
$\text{MSD}_{k>1}$. Note from (\ref{eq32}) that MSD$_{k>1}$ depends
on $\eta$ through the function $h(2/\sqrt{\eta})/\eta$, or
equivalently, $\alpha^2h(\alpha)/4$ by replacing $2/\sqrt{\eta}$
with $\alpha$. We show the derivative of $\alpha^2h(\alpha)/4$ with
respect to $\alpha$ in the right plot of Fig. \ref{Fig_3}. It is
seen that the derivative function increases rapidly beyond
$\alpha=0.8$. To maintain acceptable levels of accuracy, it is
preferable for the derivative to be bounded by a relative small
value, say, $0.5$. Then, the value of $\alpha$ should be less than
$0.8$, or $\eta\geq 6.25$. That is, the average neighborhood sizes
should be kept around 6-7 or larger.

\subsection{Convergence Rate Expression}
From (\ref{eq56}), $|\lambda_{k,m}(\mathcal{B})|$ can be expressed
as:
\begin{equation}
    |\lambda_{k,m}(\mathcal{B})|=|\lambda_k(A)|\cdot
    |1-\mu\lambda_m(R_u)\cdot s^*_{k,1:N_I}r_{k,1:N_I}|
\end{equation}
Since $|\lambda_k(A)|<|\lambda_1(A)|=1$ for $k>1$, and for
sufficiently small step-sizes, the maximum value of
$|\lambda_{k,m}(\mathcal{B})|$ (namely, $\rho(\mathcal{B})$) occurs
when $k=1$. Recall that all entries of $r_1$ and $s_1$ are positive,
which implies that $|\lambda_{1,m}(\mathcal{B})|$ increases as $m$
increases (i.e. smaller $\lambda_m(R_u)$). Then, we arrive at the
following expression for $\rho(\mathcal{B})$:
\begin{equation}\label{eq95}\boxed{
\begin{aligned}
    \rho(\mathcal{B})=|\lambda_{1,M}(\mathcal{B})|
    = 1-\mu\lambda_M(R_u)\cdot s^T_{1,1:N_I}r_{1,1:N_I}
\end{aligned}}
\end{equation}
The square of this expression determines the rate of convergence of
the ATC diffusion strategy (\ref{eq49}). Note that expression
(\ref{eq95}) satisfies Lemmas \ref{lem_1} and \ref{thm_1}. For the
uniform $A$ in (\ref{eq21}), we obtain from (\ref{eq93}),
(\ref{eq48}), and (\ref{eq95}) that
\begin{equation}\label{eq30}\boxed{
\begin{aligned}
    \rho(\mathcal{B})
    = 1-\mu\lambda_M(R_u)\cdot\frac{\sum_{l=1}^{N_I}n_l}
    {N\eta}
\end{aligned}}\quad \text{(using uniform combination matrix (\ref{eq21}))}
\end{equation}
Expression (\ref{eq30}) can be motivated intuitively by noting that
the decay of $\rho(\mathcal{B})$ will be larger as informed nodes
have higher degrees. Simulations further ahead show that expression
(\ref{eq30}) matches well with simulated results.

\subsection{Behavior of the ATC Network}
\begin{figure}
\centering
\includegraphics[width=28em]{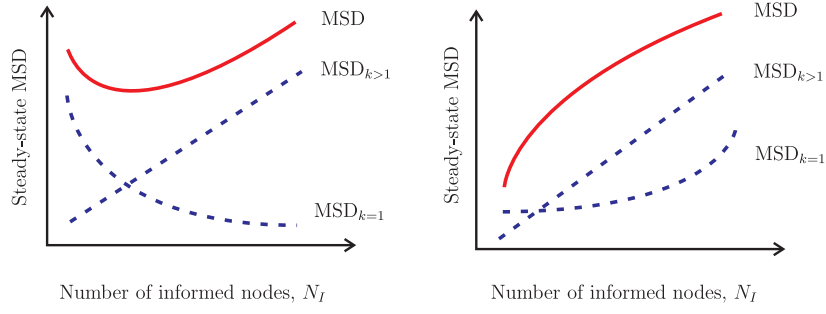}
\caption{Sketch of the behavior of the network MSD as a function of
the number of informed nodes, $N_I$, depending on whether relation
(\ref{eq29}) is satisfied (left) or not (right).} \label{Fig_7}
\end{figure}

Combining expressions (\ref{eq58}), (\ref{eq32}), and (\ref{eq30}),
we arrive at the following result for ATC diffusion networks.
\begin{thm}[Network MSD under uniform combination weights]
The ATC network (\ref{eq49}) with uniform step-sizes and regression
covariance matrices ($\mu_k=\mu$ and $R_{u,k}=R_u$) and with the
uniform combination matrix $A$ in (\ref{eq21}) has approximate
convergence rate:
\begin{equation}\label{eq43}
\begin{aligned}
    r\approx \left(1-\mu\lambda_M(R_u)\cdot\frac{\sum_{l\in\mathcal{N}_I}n_l}
    {N\eta}\right)^2
\end{aligned}
\end{equation}
and approximate network MSD:
\begin{equation}\label{eq42}
    {\rm{MSD}} \approx \underbrace{\frac{M\mu}{2N\eta}\cdot
    \frac{\sum_{l\in\mathcal{N}_I}\sigma^2_{v,l}n^2_l}
    {\sum_{l\in\mathcal{N}_I}n_l}}_{{\rm{MSD}}_{k=1}}\;+\;
    \underbrace{\frac{\mu^2{\rm{Tr}}(R_u)}{N\eta}\cdot
    h\left(\frac{2}{\sqrt{\eta}}\right)\cdot
    \sum_{l\in\mathcal{N}_I}\sigma^2_{v,l}n_l}_{{\rm{MSD}}_{k>1}}
\end{equation}
where $\eta$ and $h(\cdot)$ are defined in (\ref{eq46}) and
(\ref{eq38}), respectively.\vspace{-1cm}
\begin{flushright}
    $\blacksquare$
\end{flushright}
\end{thm}
Note that the summations in (\ref{eq43}) and (\ref{eq42}) are over
the set of informed nodes, $\mathcal{N}_I$. Expressions (\ref{eq43})
and (\ref{eq42}) reveal important information about the behavior of
the network. First, the convergence rate in (\ref{eq43}) and the
network MSD in (\ref{eq42}) depend on the network topology only
through the node degrees, $\{n_l\}$, and the network degree, $\eta$.
In general, the higher values of $\eta$ are, the slower the
convergence rate is (an undesirable effect) and the lower the
network MSD is (a desirable effect). Second, as the set of informed
nodes, $\mathcal{N}_I$, increases, we observe from (\ref{eq43}) that
the faster the rate of convergence becomes (a desirable effect).
However, as we will illustrate in simulations, the behavior of the
terms MSD$_{k=1}$ and MSD$_{k>1}$ ends up causing the network MSD
given by (\ref{eq42}) to increase (an undesirable effect) as
$\mathcal{N}_I$ increases. Figure \ref{Fig_7} illustrates the
general trend in the behavior of the network MSD and its components,
MSD$_{k=1}$ and MSD$_{k>1}$. Two scenarios are shown in the figure
corresponding to the case whether the added informed nodes satisfy
(\ref{eq29}) or not. The figure shows that depending on condition
(\ref{eq29}), the curve for MSD$_{k=1}$ can increase or decrease
with $N_I$. Nevertheless, the overall network MSD generally
increases (i.e., becomes worse) with increasing $N_I$. These facts
reveal an important trade-off between the convergence rate and the
network MSD in relation to the proportion of informed nodes. We
summarize the behavior of the ATC network in Table I and show how
the rate of convergence and the MSD respond when the parameters
$\{\eta, N_I, \text{Tr}(R_u)\}$ increase. We remark that slower
convergence rate and worse estimation correspond to increasing
values of $r$ and MSD (an undesirable effect).

\begin{table}\label{tab1}
\centering \caption{Behavior of the ATC Network in response to
increases in any of the parameters $\{\eta,N_I,{\rm{Tr}}(R_u)\}$}
\begin{tabular}{|c|c|c|c|c|}
\hline   & convergence rate $r$ (\ref{eq43}) & MSD (\ref{eq42}) &
MSD$_{k=1}$ (\ref{eq58}) & MSD$_{k>1}$ (\ref{eq30}) \\
\hline \hline $N_I\uparrow$ & faster & worse in general &
may be better or worse (see (\ref{eq29})) & worse  \\
\hline $\eta\uparrow$ & slower & better & better & better  \\
\hline $\text{Tr}(R_u)\uparrow$ & faster & worse & independent of $\text{Tr}(R_u)$ & worse  \\
\hline
\end{tabular}
\end{table}

For a proper evaluation of how the proportion of informed nodes
influences network behavior, we shall adjust the step-size parameter
such that the convergence rate remains fixed as the set of informed
nodes is enlarged and then compare the resulting network MSDs. To do
so, we set the step-size to the following normalized value:
\begin{equation}\label{eq83}
    \mu = \frac{\mu_0}{\sum_{l\in\mathcal{N}_I}n_l}
\end{equation}
for some $\mu_0>0$. Note that this choice normalizes $\mu_0$ by the
sum of the degrees of the informed nodes. In this way, the
convergence rate given by (\ref{eq43}) becomes
\begin{equation}
    r\approx \left(1-\frac{\mu_0\lambda_M(R_u)}
    {N\eta}\right)^2
\end{equation}
which is independent of the set of informed nodes. Moreover, the
network MSD in (\ref{eq42}) becomes
\begin{equation}\label{eq80}
    \text{MSD} \approx \frac{M\mu_0}{2N\eta}\cdot
    \frac{\sum_{l\in\mathcal{N}_I}\sigma^2_{v,l}n^2_l}
    {\left(\sum_{l\in\mathcal{N}_I}n_l\right)^2}+
    \frac{\mu_0^2\text{Tr}(R_u)}{N\eta}\cdot
    h\left(\frac{2}{\sqrt{\eta}}\right)\cdot
    \frac{\sum_{l\in\mathcal{N}_I}\sigma^2_{v,l}n_l}
    {\left(\sum_{l\in\mathcal{N}_I}n_l\right)^2}
\end{equation}
Using the same argument we used before in (\ref{eq79}), if we
increase the number of informed nodes by one, the first term in
(\ref{eq80}) (namely, MSD$_{k=1}$) will increase if the degree of
the added node satisfies:
\begin{equation}\label{eq81}
\begin{aligned}
    n_{N_I+1}&\geq \underbrace{2\left[\frac{\sigma^2_{v,N_I+1}
    \left(\sum_{l\in\mathcal{N}_I}n_l\right)^2}
    {\sum_{l\in\mathcal{N}_I}\sigma^2_{v,l}n^2_l}-1\right]^{-1}}_{c_1}
    \sum_{l\in\mathcal{N}_I}n_l
\end{aligned}
\end{equation}
and the second term in (\ref{eq80}) (namely, MSD$_{k>1}$) will
increase if the degree of the added node satisfies:
\begin{equation}\label{eq82}
\begin{aligned}
    n_{N_I+1}&\leq \underbrace{\left(\frac{\sigma^2_{v,N_I+1}
    \sum_{l\in\mathcal{N}_I}n_l}
    {\sum_{l\in\mathcal{N}_I}\sigma^2_{v,l}n_l}-2\right)}_{c_2}
    \sum_{l\in\mathcal{N}_I}n_l
\end{aligned}
\end{equation}
In the following, we show that there exist conditions under which
both requirements (\ref{eq81}) and (\ref{eq82}) are satisfied. That
is, when this happens and interestingly, the network MSD ends up
increasing (an undesirable effect) when we add one more informed
node in the network. In the first example, we assume that the
degrees of all nodes are the same, i.e., set $n_l = n$ for all $l$.
Then, $c_1$ and $c_2$ in (\ref{eq81})-(\ref{eq82}) become
\begin{align}
    c_1 = 2(N_I\beta -1)^{-1},\quad
    c_2 = \beta-2
\end{align}
where
\begin{equation}
    \beta = \frac{\sigma^2_{v,N_I+1}}{\sum_{l\in\mathcal{N}_I}\sigma^2_{v,l}/N_I}
\end{equation}
It can be verified that if
\begin{equation}
    \beta \geq 2+\frac{1}{N_I}
\end{equation}
(or, if the noise variance at the added node is large enough), both
(\ref{eq81}) and (\ref{eq82}) are satisfied and then the MSD will
increase (i.e., become worse). A second example is obtained by
setting the noise variances to a constant level, i.e.,
$\sigma^2_{v,l}=\sigma^2_v$ for all $l$. Then, $c_1$ and $c_2$ in
(\ref{eq81})-(\ref{eq82}) become
\begin{align}
    c_1 = 2\left[\frac{\left(\sum_{l\in\mathcal{N}_I}n_l\right)^2}
    {\sum_{l\in\mathcal{N}_I}n^2_l} -1\right]^{-1},\quad
    c_2 = -1
\end{align}
In this case, the second term in (\ref{eq80}) always decreases,
whereas the first term in (\ref{eq80}) will increase if the degree
of the added informed node is high enough. However, as the number of
informed nodes increases, the step-size in (\ref{eq83}) will become
smaller and the first term in (\ref{eq80}) becomes dominant. As a
result, the network MSD worsens if (\ref{eq81}) is satisfied, i.e.,
when the added node has large degree. \emph{These results suggest
that it is beneficial to let few highly noisy or highly connected
nodes remain uninformed and participate only in the consultation
step (\ref{eq49_2})}.

\section{Simulation Results}
We consider networks with 200 nodes. The weight vector, $w^\circ$,
is a randomly generated $5\times 1$ vector (i.e., $M=5$). The
regressor covariance matrix $R_u$ is a diagonal matrix with each
diagonal entry uniformly generated from $[0.8,1.8]$, and noise
variances are set to $\sigma^2_{v,k}=0.01$ for all $k$. The
step-size for informed nodes is set to $\mu=0.01$. Without loss of
generality, we assume that the nodes are indexed in decreasing order
of degree, i.e., $n_1\geq n_2 \geq\cdots\geq n_N$.

\begin{figure}
\centering
\includegraphics[width=34em]{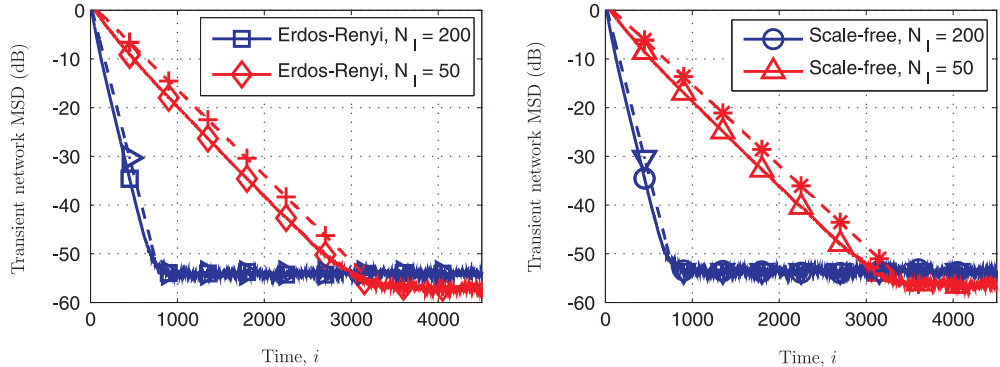}
\caption{Transient network MSD over the Erdos-Renyi (left) and
scale-free (right) networks with $200$ nodes. The dashed lines
represent the theoretical results (\ref{eq31}) and (\ref{eq89}).}
\label{Fig_1}
\end{figure}

We first verify theoretical expressions (\ref{eq31}) and
(\ref{eq89}) for the network MSD and convergence rate. Figure
\ref{Fig_1} shows the MSD over time for two network models with
parameters $p=0.02$, $m=2$, and $N_0=10$. For each network model, we
consider two cases: 200 or 50 (randomly selected) informed nodes. We
observe that when the number of informed nodes decreases, the
convergence rate increases, as expected, but interestingly, the MSD
decreases. The theoretical results are also depicted in Fig.
\ref{Fig_1}. The MSD decays at rate $r$ in (\ref{eq89}) during the
transient stage. When the MSD is lower than the steady-state MSD
value from (\ref{eq31}), the MSD stays constant at (\ref{eq31}). We
observe that the theoretical results match well with simulations.
The theoretical results (\ref{eq31}) and (\ref{eq89}) will be used
to verify the effectiveness of the approximate expressions
(\ref{eq43}) and (\ref{eq42}).

\begin{figure}
\centering
\includegraphics[width=34em]{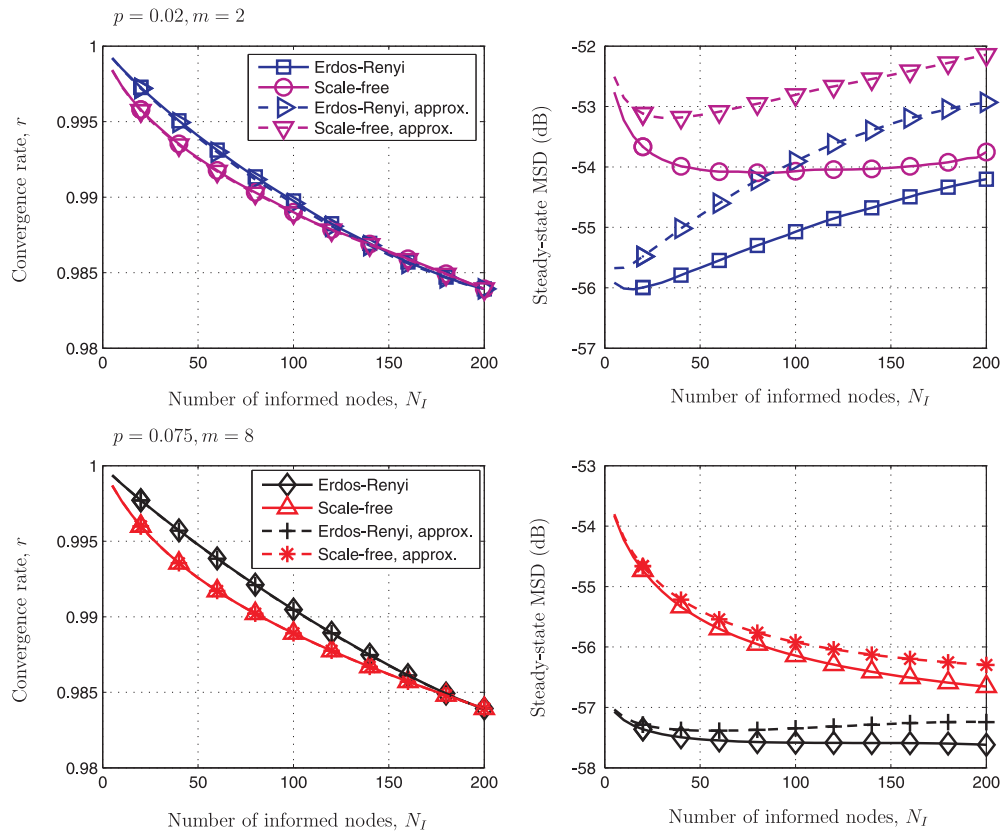}
\caption{Convergence rate (left) and steady-state MSD (right) for
Erdos-Renyi and scale-free models with the addition of informed
nodes in decreasing order of degree. The dashed lines represent
approximate expressions (\ref{eq43}) and (\ref{eq42}).}
\label{Fig_4}
\end{figure}

We examine the effect of the proportion and distribution of informed
nodes on the convergence rate and MSD of the network. We increase
the number of informed nodes from the node with the highest degree,
i.e., from node 1 to node $N$. The convergence rate and MSD are
shown in Fig. \ref{Fig_4}. For each model, we consider two possible
values of parameters: $p=0.02$ and $0.075$ in the Erdos-Renyi model
and the $m=2$ and $8$ in the scale-free model. Simulation results
are averaged over $30$ experiments. Note from (\ref{eq88}) and
(\ref{eq70}) that the two models have similar network degree. As
expected, the convergence rates decrease when we add more informed
nodes and expression (\ref{eq43}) matches well with expression
(\ref{eq89}). In addition, the convergence rates in the scale-free
model are lower in the beginning because there are some nodes with
very high degrees.

\begin{table}\label{tab2}
\centering \caption{Network degree and $|\lambda_2(A)|$ for two
network models}
\begin{tabular}{|c|c|c|c|c|}
\hline   & \multicolumn{2}{c|}{Erdos-Renyi ($p$)} &
\multicolumn{2}{c|}{Scale-free ($m$)} \\
\hline \hline Parameter ($p$ or $m$) & 0.02 & 0.075 & 2  & 8 \\
\hline $\eta$ & 5.13 & 15.83 & 4.93 & 16.33 \\
\hline $|\lambda_2(A)|$ & 0.883 & 0.503 & 0.900 & 0.495 \\
\hline
\end{tabular}
\end{table}

\begin{figure}
\centering
\includegraphics[width=38em]{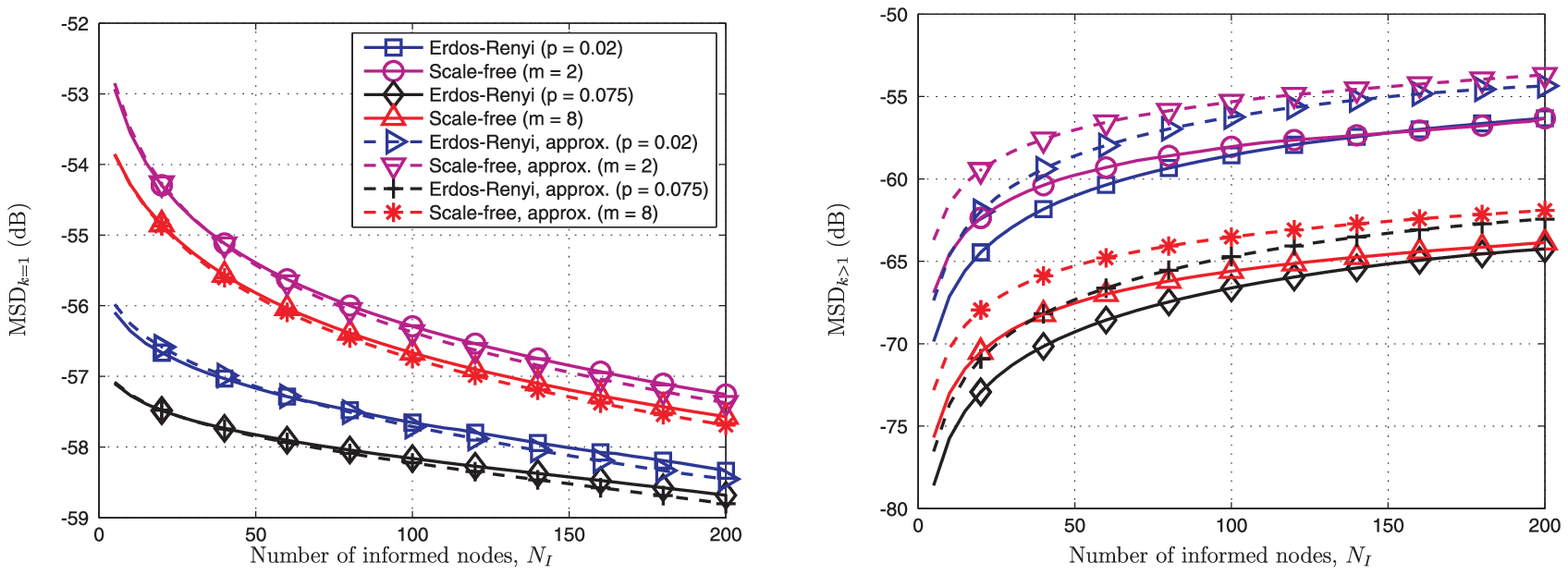}
\caption{MSD$_{k=1}$ (left) and MSD$_{k>1}$ (right) for Erdos-Renyi
and scale-free models with the addition of informed nodes in
decreasing order of degree. The dashed lines represent approximate
expressions (\ref{eq58}) and (\ref{eq32}).} \label{Fig_5}
\end{figure}

Interesting patterns are seen in the MSD behavior. We show
$\text{MSD}_{k=1}$ from (\ref{eq58}) and $\text{MSD}_{k>1}$ from
(\ref{eq32}) in Fig. \ref{Fig_5}. We observe from Fig. \ref{Fig_5}
that $\text{MSD}_{k=1}$ decreases, whereas $\text{MSD}_{k>1}$
increases with $N_I$. If two network models have similar degree, the
scale-free model will have higher values of $\text{MSD}_{k=1}$ and
$\text{MSD}_{k>1}$ than the Erdos-Renyi model, and therefore higher
values of MSD. This is because the scale-free model has higher
values of $n_l$. Since $\text{MSD}_{k=1}$ decreases and
$\text{MSD}_{k>1}$ increases, the resulting MSD in (\ref{eq42}) can
either increase or decrease. The curve of MSD depends on the values
of $\text{MSD}_{k=1}$ and $\text{MSD}_{k>1}$. We observe from Fig.
\ref{Fig_4} that in most cases, the MSD decreases when $N_I$ is
small, and then increases with $N_I$. As in the case of a
stand-alone adaptive filter, there exists a trade-off between the
convergence rate and the MSD. Interestingly, for the scale-free
model with higher values of $m$, we see from Fig. \ref{Fig_4} that
the MSD decreases with $N_I$. We also see that the approximation for
the MSD in (\ref{eq42}) matches well with expression (\ref{eq31}).

expression (\ref{eq78}) becomes
\begin{equation}
    \text{MSD} \approx \frac{M\mu\sigma^2_{v}}{2}\cdot
    \frac{\sum_{l=1}^Nn^2_l}
    {\left(\sum_{l=1}^Nn_l\right)^2}+
    \mu^2\text{Tr}(R_u)\sigma^2_v\cdot
    h\left(\frac{2}{\sqrt{\eta}}\right)
\end{equation}
From the Cauchy-Schwarz inequality and for a fixed value of $\eta$,
we know that
\begin{equation}
    \left(\sum_{l=1}^Nn_l\right)^2\leq
    N\cdot\sum_{l=1}^Nn^2_l
\end{equation}
with equality if, and only if, $n_l=\eta$ for all $l$, i.e., all
nodes have the same degree. Then, we obtain a lower bound for the
MSD:
\begin{equation}
    \text{MSD}\geq \frac{M\mu\sigma^2_v}{2N}+
    \mu^2\text{Tr}(R_u)\sigma^2_v\cdot
    h\left(\frac{2}{\sqrt{\eta}}\right)
\end{equation}
Since the nodes in the Erdos-Renyi model have similar degree, it
will achieve lower MSD than the scale-free model if all nodes are
informed.


\subsection{MSD with Fixed Convergence Rate}
\begin{figure}
\centering
\includegraphics[width=20em]{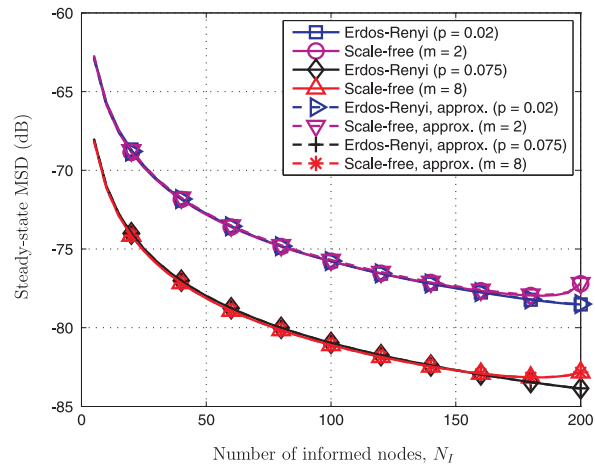}
\caption{Steady-state MSD with the deployment for node $N$ to node
$1$ for Erdos-Renyi and scale-free models. The dashed lines
represent approximate expression (\ref{eq80}).} \label{Fig_8}
\end{figure}

We vary the value of step-size as in (\ref{eq83}) with $\mu_0 = 0.1$
and show the network MSD over the number of informed nodes in Fig.
\ref{Fig_8}. To show the MSD possibly increases with $N_I$, we
reverse the order in adding informed nodes, i.e., from node $N$ to
node 1. It is interesting to note that for the scale-free model, the
MSD increases when the number of informed nodes is large. This is
because in the scale-free model, there are few nodes connected to
most nodes in the network and condition (\ref{eq81}) is satisfied.
The results suggest that in the scale-free model, we should let few
highly connected nodes remain uninformed and perform only the
consultation step (\ref{eq49_2}).

\section{Concluding Remarks}
In this paper, we derived useful expressions for the convergence
rate and mean-square performance of adaptive networks. The analysis
examines analytically how the convergence rate and mean-square
performance of the network vary with the degrees of the nodes, with
the network degree, and with the proportion of informed nodes. The
results reveal interesting and surprising patterns of behavior. The
analysis shows that there exists a trade-off between convergence
rate and mean-square performance in terms of the proportion of
informed nodes. It is not always the case that increasing the
proportion of informed nodes is beneficial.

\bibliographystyle{IEEEtran}
\bibliography{IEEEfull,refs}

\end{document}